# ISOTOPE HARVESTING AT FRIB:

*Additional opportunities for scientific discovery*

## TABLE OF CONTENTS

## Preface: The Working Group on Isotope Harvesting at FRIB

*As early as 2006, the opportunity for scientific advances with harvested isotopes from FRIB has been recognized by the nuclear science community. In 2010, the Department of Energy Office of Science, Nuclear Physics, sponsored the first Workshop on Isotope Harvesting at FRIB: convening researchers from diverse fields to discuss the scientific impact and technical feasibility of isotope harvesting. Following the initial meeting, a series of biennial workshops was organized. At the fourth workshop, at Michigan State University in 2016, the community elected to prepare a formal document to present their findings. This report is the output of the working group, drawing on contributions and discussions with a broad range of scientific experts.*

**Participants and contributors:**
Virginia Ayres, Michigan State University
Eva Birnbaum, Los Alamos National Laboratory
Georg Bollen, Michigan State University
Greg Bonito, Michigan State University
Todd Bredeweg, Los Alamos National Laboratory
Aaron Couture, Los Alamos National Laboratory
Matt Dietrich, Argonne National Laboratory
Paul Ellison, University of Wisconsin-Madison
Jonathan Engle, University of Wisconsin-Madison
Richard Ferrieri, University of Missouri
Jonathan Fitzsimmons, Brookhaven National Laboratory
Moshe Friedman, Michigan State University
Stephen Graves, University of Iowa
John Greene, Argonne National Laboratory
Suzanne Lapi, University of Alabama at Birmingham
Paul Mantica, Michigan State University
Tara Mastren, Los Alamos National Laboratory
Cecilia Martinez-Gomez, Michigan State University
David Morrissey, Michigan State University
Graham Peaslee, University of Notre Dame
J. David Robertson, University of Missouri
Nicholas Scielzo, Lawrence Livermore National Laboratory
Gregory Severin, Michigan State University
Dawn Shaughnessy, Lawrence Livermore National Laboratory
Jennifer Shusterman, Lawrence Livermore National Laboratory
Jaideep Singh, Michigan State University
Mark Stoyer, Lawrence Livermore National Laboratory
Ate Visser, Lawrence Livermore National Laboratory

**Student participants and contributors:**
E. Paige Abel, Michigan State University
Hannah Clause, Michigan State University
C. Shaun Loveless, University of Alabama at Birmingham
Sean McGuinness, University of Notre Dame
Matthew Scott, University of Missouri
Logan Sutherlin, University of Missouri
John Wilkinson, University of Notre Dame

**FRIB Technical Contacts:**
Mikael Avilov, Joe DeVore, Dali Georgobiani, Wolfgang Mittig, and Frederique Pellemoine.

## Abstract

The Facility for Rare Isotope Beams (FRIB) at Michigan State University provides a unique opportunity to access some of the nation's most specialized scientific resources: *radioisotopes*. An excess of useful radioisotopes will be formed as FRIB fulfills its basic science mission of providing rare isotope beams. In order for the FRIB beams to reach high-purity, many of the isotopes are discarded and go unused. If harvested, the unused isotopes could enable cutting-edge research for diverse applications ranging from medical therapy and diagnosis to nuclear security. Given that FRIB will have the capability to create about 80% of all possible atomic nuclei, harvesting at FRIB will provide a fast path for access to a vast array of isotopes of interest in basic and applied science investigations. To fully realize this opportunity, infrastructure investment is required to enable harvesting and purification of otherwise unused isotopes. An investment in isotope harvesting at FRIB will provide the nation with a powerful resource for development of crucial isotope applications.

## Executive Summary

The Facility for Rare Isotope Beams (FRIB) will soon be the most far-reaching rare isotope facility in the world. As part of its basic science mission it will create beams of radionuclides that have previously only existed in supernovae explosions and in the crusts of neutron stars. The normal operation of FRIB will generate a wide variety of useful radioisotopes, allowing researchers to address forefront scientific questions, and to meet diverse astrophysical and nuclear science goals that improve our understanding of the universe. Many of these radioisotopes also have applications that serve societal needs, in fields ranging from diagnosis and treatment of cancer to national security.

To take full advantage of the rare isotope inventory of FRIB, an associated radiochemistry capability is needed. Such a capability will expand the discovery footprint of FRIB through an active isotope harvesting program. The NSAC Isotopes Subcommittee, in their 2015 NSAC-Isotope Long Range Plan, recognized this need and recommended support for "*infrastructure for isotope harvesting at FRIB*":

> "*During routine operation for its nuclear physics mission, FRIB will produce a broad variety of isotopes that could be harvested synergistically without interference to the primary user. Research quantities of many of these isotopes, which are of interest to various applications including medicine, stockpile stewardship and astrophysics, are currently in short supply or have no source other than FRIB operation.*"

Central to the recommendation is realization of the opportunity for isotope harvesting at FRIB to support a strong secondary research portfolio with its complementary isotopes. Harvesting will generate even further return on the FRIB investment.

The scientific opportunity provided by FRIB, due to its ability to create rare and valuable isotopes and its physical separation scheme will be unequaled internationally. FRIB is expected to create approximately 80% of all isotopes predicted to exist, and many of these isotopes simply cannot be made anywhere else. Harvesting will make an unprecedented diversity of isotopes available to communities for pursuing otherwise inaccessible research goals across many fields, including energy, medicine, environment, and security.

A modest infrastructure investment is required to make isotope harvesting at FRIB a reality. Michigan State University (MSU) has already built an analytical radiochemistry laboratory to conduct preliminary research into isotope harvesting. Additionally, space has been set aside for high-level processing in a new technical area contiguous with the FRIB building. To establish this isotope harvesting program, an investment of approximately $10M is needed for construction, equipment, and commissioning for a radiochemical processing facility over the four-year span from FY2020-FY2023, with an anticipated operational cost of $500k/year after completion. By following a phased-build approach, the facility will be able to process a selection of desirable by-product isotopes at the start of FRIB operations.

# ISOTOPE HARVESTING AT FRIB

## Introduction

Nuclear science and the application of radioactivity benefit society in extraordinary ways. Since the discovery of radioactivity, eleven Nobel prizes have been awarded for nuclear chemistry with another five in applied radiochemistry and radiotracing [1]. Applications are widespread across multiple disciplines, including medical diagnostics and therapies; horticultural and chemical sciences; and oil and gas exploration. The benefit of using radioactivity and radioisotopes for scientific and industrial applications comes from the chemical discrimination afforded by spectroscopy and isotope tracing and the immense sensitivity of radiation detection. These attributes allow researchers to understand both natural and synthetic processes, from the global scale of oceanographic currents down to the minute scale of receptor and epitope-based identification on the surface of cells.

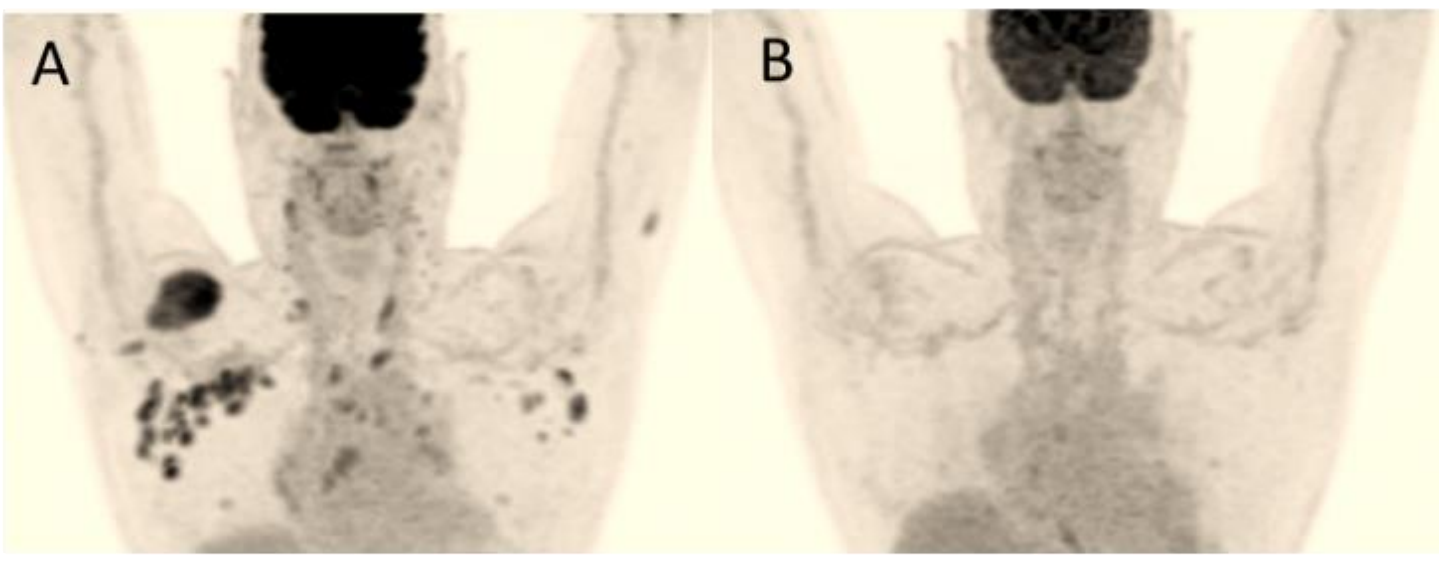

*Figure 1: [$^{18}$F]FDG PET scan of a patient with relapsed mantle cell lymphoma before (A) and after (B) two months of treatment. The extent of disease was immediately recognizable in the pre-treatment PET image, and the efficacy and completeness of treatment was easily assessed after completion of the chemotherapy course. Images courtesy of Dr. Jonathon McConathy, University of Alabama at Birmingham*

Nuclear applications and technologies improve the lives of millions of Americans each year. One of the most striking examples of the benefits of nuclear science comes from its application to medical diagnosis. Since 2005, in the US alone, the number of clinical diagnostic nuclear medicine scans has been in excess of 17 million procedures per year. Of these, the 3-dimensional tomographic technique of positron emission tomography (PET) has been used for over 1.5 million scans per year [2,3]. The primary purpose of the scans is to allow non-invasive cancer staging, which enables physicians to choose the best possible treatment for their patients. Shown in Figure 1 is an example reconstruction of a patient scanned using the tracer [$^{18}$F]fluorodeoxyglucose ([$^{18}$F]FDG), where the regions with higher contrast represent increased metabolic activity, a hallmark of many invasive cancers. The scans were obtained from the patient before and after treatment for mantle cell lymphoma. The benefit of the scans is unparalleled for treatment planning; not only does it help doctors decide what course to take, but it also gives rapid and noninvasive feedback about whether the current treatment is effective.

The technology and procedures that allow doctors to obtain and utilize PET scans are the product of public investments into nuclear science spanning many decades. Without such deep study into the nature of the atomic nucleus through the production, observation, and manipulation of isotopes, many of the major scientific advancements of the past century, like PET scanning, would not have been possible.

The recognition of the continuing and growing need for isotopes in science led to the decision by the Department of Energy (DOE) to place a subprogram within the Office of Science, Nuclear Physics that is dedicated to supporting research and development in isotope production methodologies, known as Isotope Development and Production for Research and Applications, or IDPRA. Additionally, the DOE formally manages the distribution of isotopes across the nation through the National Isotope Development Center (NIDC). Beyond this, the growing need for isotopes in industry and applied research prompted a series of studies to determine the state of isotope production nationwide. In 2015, the Nuclear Science Advisory Committee (NSAC) responded to a commission from the DOE to assess the nation's isotope needs. Following an extensive review of the current status of isotope demands and uses, the 2015 NSAC-Isotopes report was published, entitled *Meeting Isotope Needs and Capturing Opportunities for the Future* [4]. Central to the report was a description of the value of isotopes, not just in nuclear science but for the broader community as a whole.

Several examples of isotope-fueled research are highlighted in the report along three main divisions: Biology, Medicine and Pharmaceuticals; Physical Science and Engineering; and National Security and other Applications. Examples from each division are considered in the following paragraphs.

In Biology and Medicine, isotopes like $^{32}P$ and $^{14}C$ are used daily in hundreds of labs across the nation to trace biological processes in living tissues, and $^{18}F$ as per the FDG example shown in Figure 1 is used clinically year-round. Additional examples include $^{99m}Tc$ used in a vast number of tracers for medical diagnostics and research, and $^{64}Cu$ and $^{89}Zr$ for the development of new patient-specific imaging routines. The developing concept of using matched pairs of isotopes to both image and treat disease is taking hold in the growing field of theranostics with paired isotopes like $^{64}Cu/^{67}Cu$ and $^{44}Sc/^{47}Sc$. In addition, therapeutic successes with the alpha emitters $^{223}Ra$ and $^{225}Ac$ are pushing the development of other alpha emitters like $^{211}At$, and Auger emitters like $^{119}Sb$, or $^{77}Br$ and its diagnostic match, $^{76}Br$.

In Physical Science and Engineering, isotopes are widely used for industrial applications such as food and medical device irradiation, as well as mechanical wear testing. Radiothermal generators (RTGs) are used in space exploration. One important fundamental area of isotope-enabled research is in searches for an intrinsic atomic electric dipole moment (EDM). EDMs are observables that are extremely sensitive to science beyond the Standard Model of particle physics, and could help identify the root cause of the matter-antimatter imbalance in the universe – a persistent question in modern physics. There are some specific candidate nuclei that would have enhanced sensitivity to these kinds of physics, and most of them are radioactive and hard to create, like $^{225}Ra$, $^{229}Pa$, and $^{221,223}Rn$.

In National Security and Other Applications, the NSAC-I report states that "[Isotopes] have become an indispensable part of the means we use to characterize nuclear processes, and are at the heart of probes used to interrogate suspect materials." In this critical area, isotopes like $^{63}Ni$ are used nationwide in airport screening devices to ensure border security. Freight cargo entering the US is screened with transmission-source isotopes like $^{75}Se$ and $^{169}Yb$. Additionally and crucially, there is a continuing need for nuclear data for isotopes that are a part of the national stockpile stewardship program. The behavior of isotopes like $^{48}V$ and $^{88}Zr$ in an intense neutron field enables more detailed analysis of weapons-test results, and more informative post-detonation nuclear forensics.

The Facility for Rare Isotope Beams (FRIB) will be able to make all of the isotopes described in the NSAC-I report, with the potential to impact all of the above-mentioned isotope applications. This fact was recognized by the NSAC-I committee, and in the summary of their report, one of the main conclusions is that isotope harvesting at FRIB represents a significant new resource for obtaining previously unavailable and short-supply isotopes. Most importantly, the report recognizes the development of harvesting capabilities at FRIB as a high-impact infrastructure investment that deserves immediate attention as illustrated by one of the main recommendations of the NSAC-I report:

> *"Research quantities of many of these isotopes, which are of interest to various applications including medicine, stockpile stewardship and astrophysics, are currently in short supply or have no source other than FRIB operation. The technical and economic viability of this proposed capability should be developed and assessed promptly."*

The emphasis on taking advantage of FRIB's capabilities comes from the recognition that, at its core, FRIB is a high-power scientific discovery facility: providing rare isotopes to users as an electromagnetically purified beam. Importantly, an additional purification mechanism, chemical purification (implemented through a harvesting program), can operate in parallel and provide an entirely different spectrum of isotopes to researchers. Ultimately, augmenting FRIB with an isotope harvesting program will further strengthen the bond between nuclear physics and other scientific fields— bringing together scientists from many areas of expertise, ranging from nuclear security and astrophysics to horticulture and medical imaging.

*Figure 2. The FRIB Logo, depicting the scientific and societal aims of the facility: isotopes for society (yellow), nuclear physics (green), astrophysics, and fundamental symmetries (brown). Isotope harvesting at FRIB would establish additional research avenues in all four major facets of this FRIB program.*

## The unique opportunity of FRIB

The Facility for Rare Isotope Beams (FRIB) will provide the widest available range of rare isotopes for research in nuclear science and related fields. FRIB will enable fundamental nuclear science research by creating and delivering some of the most exotic nuclei in the universe. As part of normal operations, FRIB will also create many long-lived isotopes that are vital for biomedical, physical and national security applications and other branches of applied research. In fact, during routine operation of FRIB, in the process of delivering beams of exotic nuclei to the primary user of the facility, the thousands of other radionuclides created as by-products will go unused. The electromagnetic purification processes used to isolate the exotic isotope beam discard the vast majority of the co-created nuclides into a water-cooling system where they accumulate and eventually decay. Many of these long-lived radionuclides are valuable, and if they are efficiently extracted they could support multiple additional research projects without affecting the delivery of FRIB beams. New research opportunities become possible as methods are developed to extract, or "harvest", the discarded isotopes from FRIB. Exploratory research using the National Superconducting Cyclotron Laboratory (NSCL) beams has shown that isotope harvesting will be possible at FRIB with a modest investment in infrastructure and research[5–7]. This report provides an overview of the possible applications of isotopes that could be harvested at FRIB and a brief description of the steps necessary to achieve these goals.

Although not explicitly part of the project baseline, isotope harvesting at FRIB fits perfectly with the aims of the facility, as stated on the FRIB homepage:

> "FRIB will enable scientists to make discoveries about the properties of rare isotopes in order to better understand the physics of nuclei, nuclear astrophysics, fundamental interactions, and applications for society. As the next-generation accelerator for conducting rare isotope experiments, FRIB will allow scientists to advance their search for answers to fundamental questions about nuclear structure, the origin of the elements in the cosmos, and the forces that shaped the evolution of the universe." [8]

Key to the statement is the symbiotic notion of enabling discoveries in the basic sciences while also meeting the needs of society through an applied science program. One important aspect of the applied program is the creation and distribution of important and otherwise unavailable isotopes. Thus, an isotope harvesting program can provide a new and ongoing resource because FRIB was designed to make almost any isotope on the existing chart of the nuclides.

Specific areas have been identified where chemically-harvested long-lived radionuclides can be used to create short-supply and priority isotopes. These main areas are listed here and correspond to the divisions originally outlined in the 2015 NSAC-Isotopes report.

- Medicine
- Biochemistry and Materials Science
- Fundamental Symmetries and the Electric Dipole Moment
- Plant Biology and the Soil Microbiome
- Radio-Thermal Generators
- Stockpile Stewardship and Threat Reduction
- Astrophysics

As broad in scope as this list is, it is by no means exhaustive. With the development and implementation of isotope harvesting at FRIB, the nation's ability to meet its isotope needs and to respond to future demands will be greatly enhanced. Additionally, the implementation of radioisotope facilities for many sciences will also be a draw for talented students who can fill the U.S. need for a trained workforce in nuclear chemistry.

# Scientific Aims of Isotope Harvesting

## Medical applications

Since the discovery of radioactive substances, their value in medicine has been recognized. The palette of important medical isotopes is evolving in response to technological improvements in both medical instrumentation and radionuclide production methodologies. Recently, advances have come from the incorporation of radiometals and diverse radioactive halogens into molecular imaging agents including antibodies and peptides. Promising targeting results from PET scans with diagnostic radiometals and halogens are driving the development of therapeutic chemical analogs containing alpha particle and Auger electron emitters (*e.g.* [9–11]). Lanthanides with dual functionality like $^{149}$Tb (alpha and $\beta^+$) [12] motivated the CERN-ISOLDE initiative "MEDICIS", which is a European venture into isotope harvesting [13].

Much like ISOLDE, FRIB is an exceptional isotope creation machine, populating nuclides on both the proton-rich and neutron-rich sides of the chart of across all mass regions. This is an incredible opportunity to develop new medical isotopes, especially in theranostic and matched pairs, alpha emitters, and Auger electron emitters.

One clear example of the importance of FRIB harvesting for medicine is the isotope Astatine-211 ($^{211}$At). $^{211}$At is a high-priority radionuclide for medical research and clinical therapy. With a 7-h half-life, it is one of the few alpha-emitting radionuclides with an appropriate lifetime for clinical medical use that is not burdened by an extended decay chain (see Figure 3). While some production sites are currently operational (e.g. University of Washington, University of Pennsylvania, and Duke University [10,14,15]), the moderate half-life of $^{211}$At constrains the distance the isotope can be distributed. The limited number of production sites coupled with the limited distribution time leads to a severe restriction in patient access to this potentially life-saving isotope.

A recent meeting of the DOE-organized *University Network for Accelerator Production of Isotopes* (March 2017, Germantown MD) focused on tackling the problem of the $^{211}$At shortage. During the meeting, the advantages of isotope harvesting from FRIB were evident. At FRIB during $^{238}$U fragmentation, an $^{211}$At precursor, the generator parent $^{211}$Rn, is created in high quantity. The amount of $^{211}$Rn that will be created will enable extraction of $^{211}$At in comparable quantities to the largest reported US cyclotron productions. Furthermore, there is an added advantage to using $^{211}$Rn for the creation of $^{211}$At via a generator, as the longer half-life (15 h) of $^{211}$Rn allows shipment over longer distances and multiple extractions from a single generator.

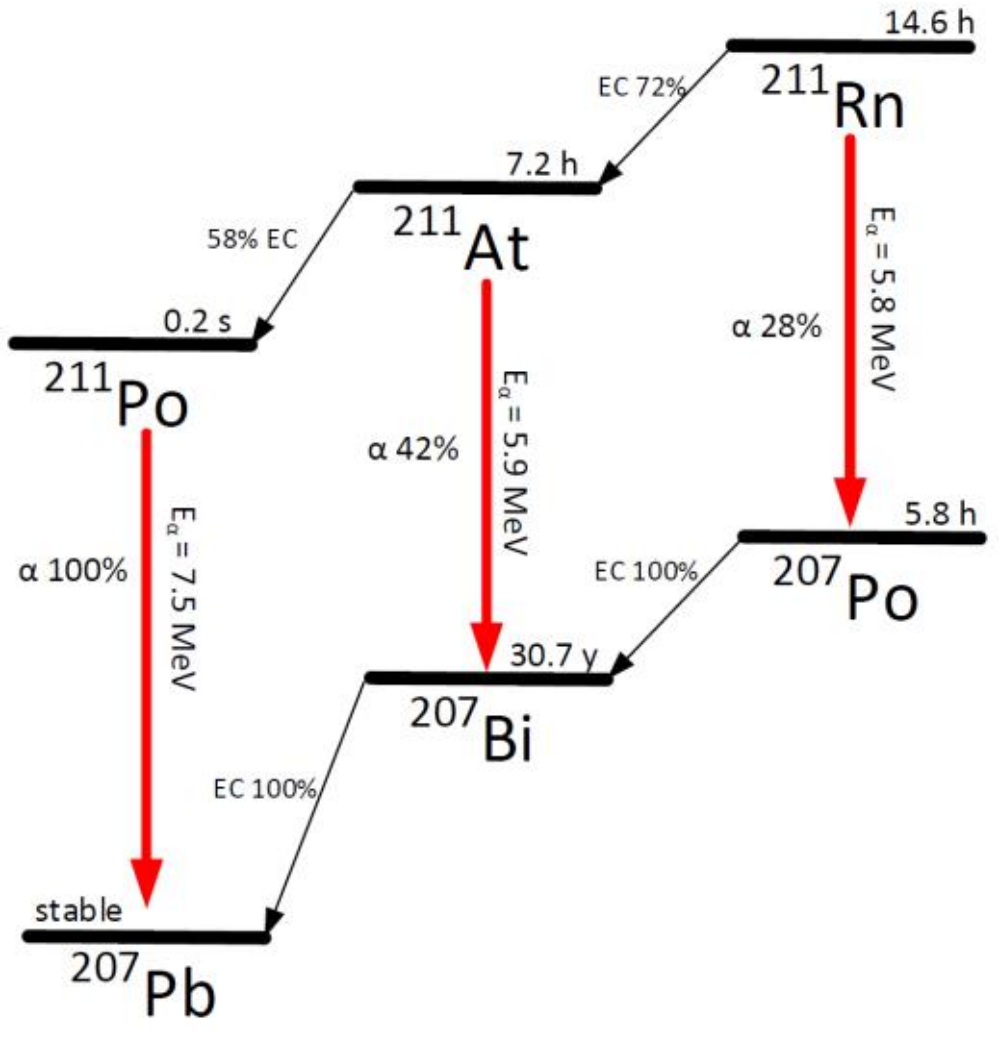

*Figure 3: A simplified decay scheme showing the important transitions for the decay of $^{211}$Rn, $^{211}$At, and their daughters*

As a result, isotope harvesting at FRIB has the potential to impact the lives of patients by providing access to a key medical isotope.

Isotopes harvested from FRIB also have significant medical research applications. A wide-reaching and exciting isotope for medical research, both in physiology and diagnosis, is $^{52}$Fe. During irradiations at FRIB using a $^{58}$Ni beam, approximately $10^{11}$ $^{52}$Fe nuclei will be formed in the FRIB beam dump every second, reaching multiple curies (Ci's) of activity in the steady-state. $^{52}$Fe is extremely important for two reasons: first because it is the only viable iron isotope for PET imaging, and second because it decays to the positron emitting short-lived manganese isotope $^{52m}$Mn. A reliable source of $^{52}$Fe could have impact for direct application, as well as for $^{52}$Fe/$^{52m}$Mn generators. A readily available $^{52}$Fe/$^{52m}$Mn generator could make key contributions to many fields, including for oncology, neurophysiology, and diabetes research owing to the critical role of manganese in biological systems [16,17].

Other examples of medically-relevant isotopes that can be harvested from FRIB in high yield are given in Appendix 1, for example: $^{44}$Ti as a parent for the positron emitter $^{44}$Sc, and $^{47}$Ca as a parent for $^{47}$Sc, the theranostic match to $^{44}$Sc; $^{76}$Kr as a parent to positron emitter $^{76}$Br and as a chemically analogous generator system to $^{211}$Rn/$^{211}$At; the Auger emitter $^{119}$Sb from its parent $^{119}$Te; and $^{72}$Se as a parent to the positron emitter $^{72}$As. These are just a few examples of the many possibilities for isotope harvesting from FRIB.

It is also important to note that the field of medical imaging is rapidly developing, and new technologies are constantly emerging. One example is in a novel modality called Polarized Nuclear Imaging (PNI) [18], described in the sidebar. As exciting new imaging technologies like PNI emerge, the isotope demands of the medical community will change. Since FRIB will make almost every isotope imaginable, harvesting from FRIB will be one of the best ways to promptly access new and important isotopes for medical research.

### Polarized Nuclear Imaging: An emerging technology with a need for isotopes

A new imaging technology, termed Polarized Nuclear Imaging (PNI), was recently unveiled in an research letter to *Nature* by the group of Gordon Cates at the University of Virginia [18]. Professor Cates and coworkers successfully demonstrated that gamma decay anisotropy from polarized nuclei could be magnetically manipulated to create a 3D image, in a manner similar to magnetic resonance imaging (MRI) [18]. The major breakthrough of PNI is to combine the extreme spatial resolution of MRI with the detection sensitivity of gamma cameras. If this technology is developed to its full potential, the major limitations of both MRI (sensitivity) and PET (resolution) will be overcome in one modality.

One requirement for successful development of PNI is to have a wide selection of short-lived polarizable gamma emitters readily available. FRIB is uniquely situated to provide access to these nuclei, and its role in developing the PNI technology is recognized in Professor Cates' Nature article, noting that even in the current stage of development, ahead of medical application:

> "… the possibilities are numerous, particularly with the ongoing construction of the US Department of Energy Facility for Rare Isotope Beams"

## Biochemistry and materials: probing hyperfine interactions with exotic nuclei

The medical uses of radioactive nuclei described above are based on organism-scale interactions between radio-labeled pharmaceuticals and organs, tissues, and even cells. For interactions on the atomic scale, there are two extremely powerful and well-established rare-isotope techniques: Mossbauer Spectroscopy, and Perturbed Angular Correlations (PAC). These important tools allow researchers to explore the interactions between nuclei and their immediate atomic surroundings.

When a nucleus is influenced by magnetic and electric fields (either arising from the chemical environment or from external sources), two important interactions occur. First, the energy levels of the nucleus' excited states shift very slightly. And second, the nucleus begins to precess in a well-defined pattern. With the first effect, even the biggest energy shifts are minute-- on the order of nano-electron Volts ($10^{-9}$ eV). Amazingly, these tiny changes are observable using Mossbauer spectroscopy, a technique that utilizes resonant absorption of gamma rays and the Doppler Effect to measure energy-level splitting and shifts with extremely high precision. By understanding the changes to nuclear energy levels, attributes of the local chemical environment can be inferred. The second effect, the spin precession of the nucleus, is observable by PAC, a technique that deduces precession rates by measuring the spatial and temporal relationship between correlated gamma rays. Since the rate of precession is highly dependent on the magnitude and shape of the local fields, by observing the precession, a wealth of chemical knowledge becomes available (e.g. rates of ligand exchange at metal centers of enzymes [19]).

Both of these valuable techniques rely upon a very small subset of radioactive nuclei, many of which are extremely limited in availability. Mossbauer spectroscopy requires nuclei with low energy excited states that are populated by the decay of a long-lived parent isotope. PAC requires nuclei that decay through a gamma-ray cascade, passing through excited-state isomers with lifetimes comparable to the nuclear precession frequency. A recent review of the use of PAC in time resolved enzyme studies supplies a list of PAC isotopes, and states that "The major limitation of the technique is availability and production of radioisotopes with appropriate properties for PAC spectroscopy" [20]. This statement highlights the need and opportunity for isotope harvesting at FRIB. Appropriate isotopes for both PAC and Mossbauer spectroscopy will be formed in the FRIB beam dump continuously during normal operations— once these isotopes are extracted, there are countless scientific questions to tackle.

Of the many strategic areas in which harvested isotopes will play a key part is in understanding the role of metal ions in enzymes in their native state. The recent breakthrough discovery of biologically active enzymatic lanthanide ions opens a whole field of research where PAC and Mossbauer isotopes can answer basic science questions (see sidebar). Here isotopes like $^{141}$Ce, $^{145}$Pm, and $^{147}$Eu for Mossbauer, and $^{140}$La and $^{149}$Eu for PAC can be used to determine the coordination environment of the lanthanides in newly discovered proteins. Understanding the catalytic role of lanthanides and other metals in enzymes will not only improve our understanding of natural processes, but may also lead to the development of novel mimetic catalysts, or engineered enzymes that will impact global resource use.

In addition to investigating biological catalysis, hyperfine studies with exotic nuclei can also be applied to purely synthetic catalytic systems. For example, there are many novel surface- and nano-catalytic structures under development for making energy-intensive processes, like the Fischer-Tropsch Synthesis, more efficient [21]. As successful structures are discovered, PAC and Mossbauer spectroscopy will aid in discerning the reactive pathways and reactive species, which will lead to better uses of energy and material resources (e.g. [21–24]). These investigations will be extremely valuable to the development of heterogeneous noble-metal catalysts like rhodium (A=100 PAC) and ruthenium (A=99 Mossbauer) where isotope availability has been a major limitation to ongoing research. Even the most common Mossbauer and PAC isotopes $^{57}$Co, and $^{111}$In, are in short supply, and FRIB harvesting will dramatically increase their availability. (See Appendix 1 for production rates).

PAC and Mossbauer spectroscopy are just two examples of techniques where isotopes from FRIB facilitate scientific discovery and advances in biochemistry and materials science. From both the basic science and application-driven sides of research and development, isotope harvesting at FRIB will provide a critical supply of crucial radionuclides.

### Lanthanides, Life, and Natural Resources

Up until five years ago, the *f*-block rare-earth elements, also known as the lanthanides, had no known biological role. Then a surprising discovery was made: a certain class of single-carbon utilizing bacteria, methylotrophs, could incorporate the lighter lanthanides in the place of calcium in a key enzyme for methanol oxidation. Amazingly, not only were the bacteria able to use lanthanides, but they were actually thriving with them [41]. It turns out that in the presence of lanthanides, the methylotrophs create a second, rare-earth-specific enzyme that is ten times more efficient than the calcium containing enzyme. Additionally, the bacteria became avid lanthanide accumulators, stripping all available rare earths from their surroundings [42].

While this finding is interesting from a scientific point of view, the applied implications are immense. First, because the new enzyme activity, if understood mechanistically, could be manipulated to catalytically convert single-carbon compounds into commodity products. Second, the way in which the bacteria sequester lanthanides from the environment, via siderophore action, could be utilized for highly valuable lanthanide recovery [43].

Mossbauer spectroscopy and PAC, with harvested isotopes from FRIB, will be important tools to discover both how the enzymes work, and how the siderophores bind the lanthanides. One key point will be to discover the differences in coordination chemistry in the enzymes and siderophores for light versus heavy lanthanides.

## Trace-nutrient transport in plants, soil, and the microbiome

Another one of the exciting opportunities that isotope harvesting from FRIB offers is to conduct tracer studies within plants and the soil microbiome. Just below the surface of the soil, complex systems of fungi, bacteria, and plants are in constant flux, with the organisms sharing and competing for valuable short-supply resources. In fact, the recent renaissance of discoveries into the role of microorganisms in the human gut extends directly to the soil; life as we know it is not possible without the cooperation of many diverse forms of life.

Inter-kingdom interactions are essential in plant and soil systems, and new tools and techniques are needed for live imaging of biological interaction and biochemical processes in plants and soils. Radiotracers offer unique opportunities to image functional processes below-ground, as well as interactions within plants and between microorganisms in soils, and plant-symbioses. In addition to acquiring macro- and micronutrients from their soil environment, plants exude various compounds from their roots as a means for chemical communication to attract or repel microbial symbionts. Chemical interactions between roots and microbes can both directly and indirectly affect systemic plant physiology. Thanks to the adaptability of nuclear imaging, non-invasive imaging techniques like PET and SPECT can now be used to follow metabolic processes that regulate these complex interactions between members of the plant and soil microbiome (example given in figure 5).

As researchers begin to understand the relationships between the soil microbiome and higher plants, vital deficiencies can begin to be addressed. For example, strategies to support biofuel cropping systems on marginalized land could be developed through inoculation protocols that improve

### $^{48}$V, $^{90}$Mo and the global nitrogen cycle

Enzymatic nitrogen fixation is one of the most important natural processes on the planet. Higher plants, such as food crops and trees, require ammonium to flourish; however, these organisms lack the ability to convert the highly stable $N_2$ molecule to useful ammonium on their own. Therefore, these vital plants have grown to depend on symbiosis with other organisms such as the nitrogen fixing soil bacterium *Azotobacter vinelandii* to acquire the essential reduced nitrogen.

Molybdenum and vanadium have surprising roles in the relationship between *A. vinelandii* and plant nutrition— they are the centers of key metalloenzymes used by the bacterium to fix nitrogen. Due to their importance, trees that benefit from *A. vinelandii* have co-evolved to slowly sequester vanadium and molybdenum from the soil, and redistribute the metals to their leaves. When the leaves fall and form a decomposing leaf litter, the metals become readily available to the bacteria [46,47].

The cycle of metal transport is just one example of the interdependence of the organisms of the soil microbiome: a relationship that is only recently becoming appreciated, and is far from being understood. Radioisotopes like $^{90}$Mo and $^{48}$V will allow researchers to trace the transport and use of key micronutrients. This will reveal the key constituents to healthy soils; leading to more efficient use of fertilizers and more sustainable crop management through a holistic approach to addressing soil deficiencies.

plant nutrient access and plant fitness. Rare isotopes, when harvested from FRIB, will be particularly useful for radiotracer experiments that probe the role potential inoculants play in the soil. The results of which can be used to increase the bioavailability of key minerals like molybdenum, zinc, and manganese. These metals are important cofactors in plant enzymes that synthesize essential aromatic amino acids, without which, many plants cannot thrive. Additionally, the distribution and turnover of trace-metals that aid in nitrogen fixation in soil and plant microbiomes will provide insights that could lead to more efficient use of fertilizers (e.g. molybdenum and vanadium as discussed in the sidebar).

FRIB provides a unique opportunity for transition metal studies because it creates many radiometals simultaneously as a consequence of normal operation. One extremely useful extension to existing studies is in PET imaging with hard-to-produce isotopes like $^{52}$Fe, $^{52m}$Mn, $^{90}$Mo, $^{62}$Zn, $^{57}$Ni, and $^{48}$V. Such non-invasive radioisotope methods can play a vital role in studying the processes responsible for essential mineral transport, plant and microbiome immunity and competition, and responses to changing environmental conditions. Advances in the understanding of these aspects of plant physiology may lead to many breakthroughs, including new advances in food production, improved nutritional value in crops, and sustainable biofuel generation. The internationally renowned Michigan State University College of Agriculture is very well positioned to add to its leadership portfolio through ease of collaboration with FRIB.

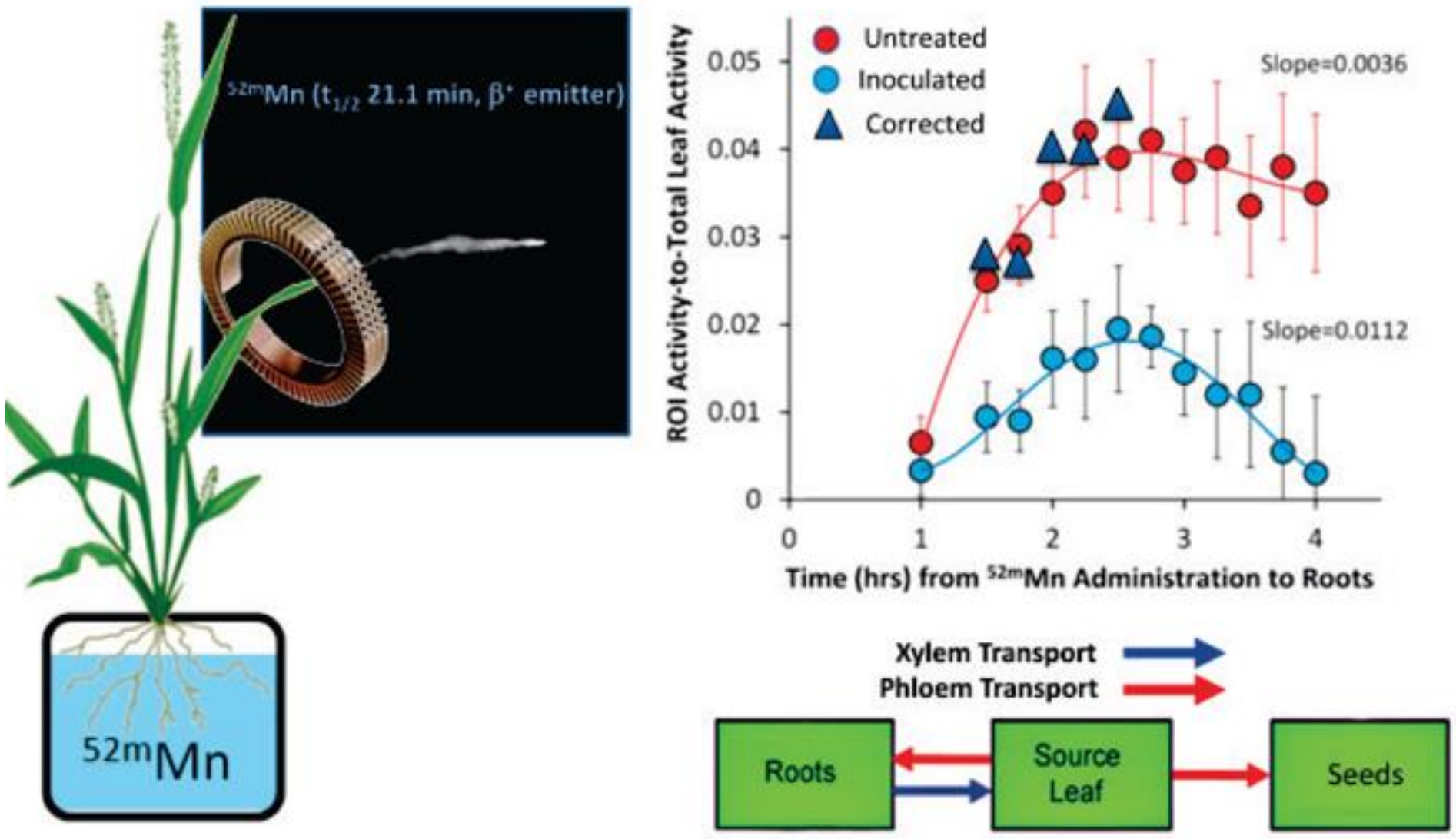

*Figure 5: A graphical depiction of an experiment exploring plant rhizosphere interactions. Plants that have been inoculated with different rhizosphere microbes exhibit different micronutrient uptake rates, which are quantified with PET detectors. One important impact of these studies is in the field of phytostimulation, where a better understanding of the complex interactions in the soil microbiome will lead to more efficient nutrient use, and overall healthier plants. With access to additional radiotracers from FRIB, the transport of many additional micronutrients can be explored. Data courtesy of Prof. Richard A. Ferrieri, Missouri University Research Reactor Center.*

## The electric dipole moment and physics beyond the standard model

Isotope harvesting is also a very important venture for expanding the nuclear physics impact of FRIB. One exciting area where harvested isotopes will play a large role is in physics-beyond-the-standard-model (BSM) experiments such as the search for the atomic electric dipole moment (EDM).

EDM searches are motivated by the persistent question: *Why is there almost no antimatter in the Universe?* The answer may be related to the existence of forces between subatomic particles that violate certain fundamental symmetries. Although the standard model of particle physics already incorporates some sources of fundamental symmetry violations discovered decades ago, it has been shown that the degree of this known amount of symmetry violation is not sufficient to explain the absence of antimatter. At the same time, BSM physics, such as supersymmetry, naturally predict additional sources of fundamental symmetry violations.

An unambiguous signature of the requisite symmetry violations would be the existence of a nonzero electric dipole moment (EDM). Calculated SM EDMs are immeasurably small for all planned upcoming experimental approaches, so any observation of an EDM in the foreseeable future would be a discovery of BSM physics. These smaller scale and often-times table-top experiments have complementary sensitivity to the Large Hadron Collider (LHC) while at the same time having sensitivity to BSM physics at the TeV-energy scale, which is beyond the reach of planned accelerator-based searches. Based on this unique and clean discovery potential, there is a world-wide effort to search for EDMs in ultracold neutrons; polar diatomic molecules; and diamagnetic atoms, each of which are sensitive to different combinations of new sources of symmetry violations. The most stringent constraints for new sources of violations originating from within the nuclear medium are mostly derived from the atomic EDM limit of Mercury-199 ($^{199}$Hg), which has a nearly spherical nucleus. Isotopes with highly deformed

### Beyond Standard: ANL's search for an EDM in $^{225}$Ra

What is the origin of the visible matter in the Universe? More specifically, why is there more matter than antimatter in the observable Universe? The answer to these questions may be visible in tiny variations to the atomic structure of exotic atoms like radium-225. At Argonne National Laboratory, a research team lead by Matt Dietrich is probing isolated $^{225}$Ra atoms to determine whether or not its deformed nucleus also distorts the distribution of charge within the surrounding electron cloud. If so, these atoms simultaneously violate both Parity and Charge Symmetries, and thereby provide a possible explanation for the observed dominance of matter over antimatter in the visible universe.

The experiment at ANL applies state-of-the-art techniques in atomic physics to answer this important nuclear physics question. Currently, Dr. Dietrich's team uses $^{225}$Ra from a legacy $^{229}$Th generator through the NIDC. However, at FRIB $^{225}$Ra will be produced directly by the $^{238}$U beam at a rate of about $10^9$ particles per second. With a steady supply of $^{225}$Ra, the ANL researchers and collaborators at other institutions can fine tune their equipment and cut down on statistical uncertainty. In the future, these developments could lead to an even more sensitive EDM search using radium molecules [44,45].

pear shaped nuclei such as Radium-225 and Protactinium-229 have an enhanced sensitivity, and are expected to have atomic EDMs that are ~$10^3$ and ~$10^5$ respectively larger than for $^{199}$Hg [25] (depicted in Figure 4). Motivated by this discovery potential, researchers at Argonne National Laboratory and their collaborators are actively searching for answers [26,27] to these long-standing questions, and isotopes from FRIB could play a key role (see sidebar).

As FRIB begins making irradiations with its $^{238}$U beam, large amounts of $^{225}$Ra, $^{229}$Pa, $^{221}$Rn and $^{223}$Rn will all be created. In the case of the shorter-lived radon isotopes, harvesting from the gas-phase and membrane contactors at FRIB may be the only viable place in the world to access the quantities needed to perform EDM experiments. $^{229}$Pa, likewise is not available from any generators or common production facilities within the US. Discoveries like the observation of an EDM will only be made from investment in a resource-mining strategy like isotope harvesting.

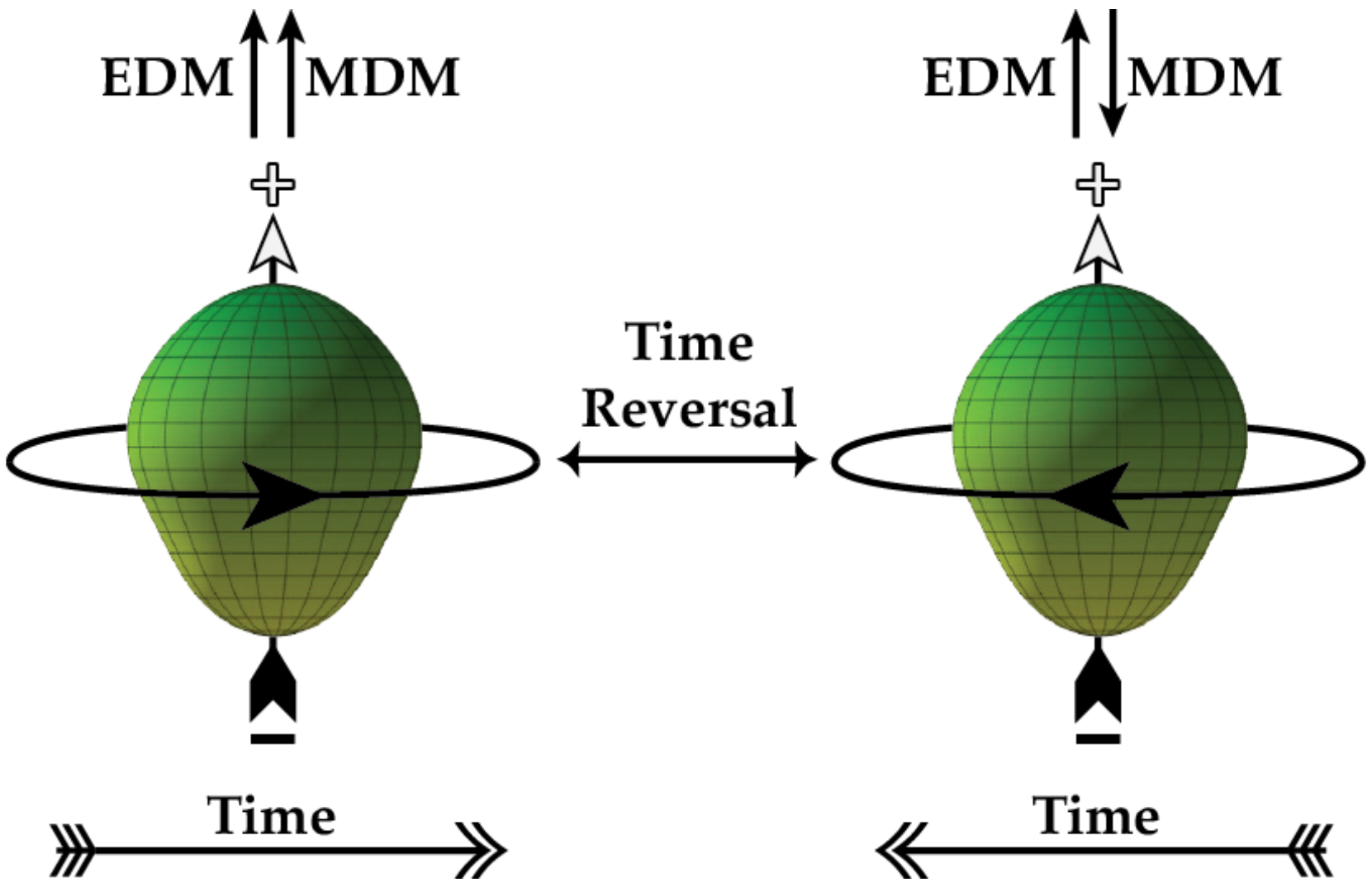

*Figure 4: Pear-shaped nuclei with both non-zero quadrupole and octupole moments have high discovery potential for EDMs because of their distorted nuclear shapes. Under the time-reversal operation, the interaction of a magnetic dipole moment (MDM) with a magnetic field will remain symmetric, whereas the interaction of an EDM with an electric field will not. Therefore, a non-zero EDM implies that the principle of microscopic reversibility of physical interactions does not apply in all cases. Observation of this type of asymmetry could help to explain the nearly complete absence of antimatter in the visible universe. Image courtesy of Prof. Jaideep Singh, Michigan State University*

## Radio-thermal generators

Beyond radiotracing, isotopes from FRIB are also in high-demand as the workhorses for micro-electro-mechanical systems (MEMS). MEMS have been developed for use as thermal, magnetic and optical sensors and actuators; as micro chemical analysis systems; as wireless communication systems; and as biomedical devices. Not surprisingly, owing to their importance in modern society, MEMS are designated as a transformational technology by the Defense Advanced Research Projects Agency (DARPA).

The ability to employ these transformational technology systems as portable, stand-alone devices in both normal and extreme environments depends, however, upon the development of power sources compatible with the MEMS technology. This is where harvested isotopes play an important role in radioisotope micro power sources (RIMS). RIMSs or "nuclear batteries" are top prospects for MEMS because they can operate for extended periods of time and in extreme environments. More importantly, because the energy change "per event" in radioactive decay is $10^4$ to $10^6$ times greater than that of a chemical reaction, the energy density (J/kg) of radioactive material is approximately $10^6$ times greater than that of lithium ion batteries. Thus, RIMS hold great potential especially when the MEMS systems are deployed in extreme and/or inaccessible environments requiring long life without recharging or refueling.

Several FRIB isotopes are strong candidates for producing a RIMS source, and are listed in table 1. For example, $^{63}$Ni could be used in RIMS with thermoelectric conversion systems; $^{147}$Pm is an ideal candidate for betavoltaic RIMS; and $^{148}$Gd is the ideal alpha emitter for liquid-semiconductor nuclear batteries or for mini-radiothermal generator systems. These valuable isotopes are all made as a consequence of normal operation at FRIB, where some creation rates will be in large excess to the currently available inventories. As an example, $^{148}$Gd is only available domestically in very small (µCi) amounts. While not enough to move into large scale device production, the 500 µCi amounts of $^{148}$Gd that will be created weekly in the primary beam dump at FRIB would be of great value in demonstrating these new, transformational technologies.

Table 1. Some candidate nuclides for radioisotope micropower sources

| **Radioisotope** | **$E_{avg}$ (keV)** | **$t_{1/2}$ (years)** | **Power (nW/mCi)** | **FRIB (mCi/wk)*** |
|---|---|---|---|---|
| $^{63}$Ni | 17.4 | 100.1 | 13 | 5.6 |
| $^{147}$Pm | 61.8 | 2.6 | 43 | 26 |
| $^{204}$Tl | 245.0 | 3.8 | 67 | 1.4 |
| $^{148}$Gd | 3200 | 75 | 1887 | 0.5 |

*see Appendix 1

## Stewardship science applications

National security is another important field where harvested isotopes from FRIB will provide a unique resource. In particular, the national stockpile stewardship program (SSP) can use FRIB isotopes to improve nuclear data for radiochemical monitor reactions in nuclear devices. The success of the SSP is vital to provide high confidence in the safety, security, reliability, and effectiveness of the U.S. stockpile.

### Radioactive Targets of Harvested $^{88}Zr$

Jennifer Shusterman, Dawn Shaughnessy, Mark Stoyer, and Nicholas Scielzo at Lawrence Livermore National Laboratory are leading efforts to measure the $^{88}Zr(n,\gamma)^{89}Zr$ cross- section in close collaboration with researchers from multiple universities across the US. Separation development and analogous neutron irradiation have been performed on samples of $^{88}Zr$ produced at cyclotrons, and are scheduled for harvested material from FRIB's predecessor, the NSCL. Separations to isolate the $^{88}Zr$ were developed to produce a pure $^{88}Zr$ target for neutron irradiation at the University of Missouri Research Reactor.

Isotope harvesting efforts lend well to student participation and will involve undergraduate and graduate students as well as postdoctoral researchers. The collaboration between LLNL and several Universities on the SSP efforts will provide an opportunity for students to visit and gain experience with projects in a national laboratory environment.

FRIB harvesting enters the spotlight in SSP because many isotopes are needed for experiments to determine the likelihood, or cross-section, of neutron-induced reactions on the radiochemical monitor nuclei. For instance, the $^{88}Zr(n,\gamma)^{89}Zr$ cross section is relevant to understanding the reaction network that is used to interpret nuclear device performance.

Currently, direct experimental access to neutron-capture reactions on radioactive nuclei is limited, and theoretical calculations of these cross sections often have uncertainties on the order of 100% or larger. For neutron-deficient nuclei, the extraction of isotopes deposited in the FRIB beam dump (or other collection locations) have the potential to provide a valuable route to gather large samples of long-lived radioisotopes. Despite being created in a veritable 'soup' of nuclei, there is expected to be a lower mass of neighboring nuclei (sub-microgram quantities) compared to the excess target material present when producing isotopes using more traditional methods with light-ion accelerators (milligram quantities). This allows for potentially higher-purity targets. Isotopes of interest can accumulate any time the facility is running, eliminating the need to obtain dedicated beam times at FRIB to access these SSP relevant nuclei. If higher isotopic purity of the harvested sample is required than is possible to obtain from aqueous collection, alternate harvesting locations at FRIB such as activated beam stops or even a dedicated experiment could be pursued.

Once FRIB is online, 1-10 μg quantities of many long-lived radionuclides should be accessible through isotope harvesting techniques and these yields offer the opportunity to perform direct cross-section measurements on harder-to-access radioactive nuclei. Example long-lived isotopes of interest are presented in Table 2 as well as the approximate activity of $10^{16}$ atoms of each of the radionuclides. To make cross section measurements, the amount of a radionuclide required depends on the properties of that radionuclide, its radiochemical purity, and the facility at which the measurement is to be made. It is expected that most of these targets will require between $10^{13}$ and $10^{15}$ atom-per-target for a measurement. These samples would have to be chemically purified and prepared as a target for irradiation with intense neutron fluxes at a reactor or a neutron-beam facility, generating inter-institutional collaboration links and opportunities.

Table 2. Some examples of radioactive isotopes of interest for the Stockpile Stewardship Program. Target activities assume a sample of $10^{14}$ atoms is required for a cross-section measurement, however, this value will vary depending on the specific radionuclide, radiochemical purity, and experimental facility.

| **Isotope** | **Activity/ target (mCi)** | **FRIB rate (mCi/wk)*** |
|---|---|---|
| $^{88}Zr$ | 20 | 630 |
| $^{48}V$ | 1400 | 2600 (direct), 55 (from $^{48}Cr$) |
| $^{168}Tm$ | 30 | 58 |
| $^{149}Eu$ | 30 | 83 (from $^{149}Gd$) |
| $^{150}Eu$ | 3 | 0.06 |
| $^{88}Y$ | 20 | 240 (from $^{88}Zr$) |
| $^{173}Lu$ | 10 | 0.65 |

*see Appendix 1

## Nuclear Astrophysics

The field of nuclear astrophysics also stands to benefit from isotope harvesting efforts at FRIB. Just recently, a similar European harvesting project called ERAWAST (exotic nuclides from accelerator waste), enabled a series of astrophysical measurements on the exotic nuclei $^{60}Fe$, $^{53}Mn$, and $^{44}Ti$ (amongst many others, see sidebar) [28,29]. These kinds of measurements were extremely important for nuclear astrophysicists striving to discover how visible matter came into being and how it evolves.

As proposed in the National Research Council's 2013 Nuclear Physics review [30], and addressed as part of the 2015 NSAC Long Range Plan [31], the key to understanding astrophysical observations lies in understanding the origins of the elements as well as the life and death of stars. Here unstable isotopes play a critical role, both in the cosmos and in the laboratory. This is because with facilities like FRIB, the nuclear reactions that create the elements we observe in the cosmos can be recreated in a lab setting. By observing both the controlled reactions, and those in the cosmos, the combined data can be used to understand, diagnose, and constrain astrophysical environment models.

Harvesting isotopes from FRIB will play a role due to the nature of some of the ongoing investigations into astrophysical reactions that lead to heavy elements. The elements heavier than iron are made primarily through neutron capture scenarios, with the *slow* and *rapid* neutron capture processes (*s* and *r* processes, respectively) each accounting for approximately half of the observed abundance of the heavy elements [32]. The *r* process is expected to take place in explosive scenarios on a time scale of seconds to tens of seconds. While fast and reaccelerated beam FRIB experiments will provide a wealth of new information to inform *r*-process nucleosynthesis, observations and experiments for the *s*-process involve interactions on an entirely different timescale. There, the critical unstable isotopes are the *s*-process branch-point isotopes, which have half-lives ranging from hundreds of days to tens of years. In order to understand the behavior of these isotopes, samples need to be collected for long periods of time in order to accumulate enough material to recreate relevant nuclear reaction scenarios. Harvesting is a perfect fit for these experiments, because it is the only realistic approach that allows isotopes to accumulate for several years without demanding any dedicated beam-time.

One important harvesting target for studying the s-process is $^{85}$Kr. During the s-process, $^{85}$Kr can either beta decay to form $^{85}$Rb, or it can capture a neutron, creating $^{86}$Kr. $^{86}$Kr is important because it is mostly produced in the s-process [33], and it is close to the end of the weak s-process component (massive stars) around mass A=90. The main component (AGB stars), produces significant amounts of $^{86}$Kr and the freshly synthesized material is implanted into presolar grains [34,35]. The ($^{86}$Kr/$^{82}$Kr) ratio is strongly affected by the s-process branching point at $^{85}$Kr, but the recommended $^{85}$Kr(n,γ)$^{86}$Kr cross-section at stellar energies is based only on theoretical estimates and has an uncertainty of almost a factor of two [36]. Changing the cross-section by a factor of two in the M = 1.5 $M_{\odot}$, [Fe/H] = -0.30, standard AGB star case, the predicted $^{86}$Kr/$^{82}$Kr ratio varies by 80% [37]. It is therefore very difficult to obtain a well-established estimate for the $^{86}$Kr/$^{82}$Kr ratio from AGB models, and at least a factor of two of uncertainty must be

## ERAWAST- European Isotope Harvesting, and Nuclear Astrophysics

In 2006 Dr. Dorothea Schumann of the Paul Scherrer Institute (PSI) proposed a novel use for aged accelerator components at PSI's high energy beam facility: to mine them for valuable radioisotopes [28]. Soon after, the project, termed ERAWAST, led by Dr. Schumann undertook harvesting long-lived radionuclides from one of PSI's copper beam-stops [29].

Inside of the beam-stop was one of the most sought-after radionuclides for nuclear astrophysics, $^{60}$Fe. Outside of the laboratory, this 2.6 My half-life isotope of iron is formed as a result of extreme cosmic events, such as supernovae. Because it can be observed both in space and in meteorite samples, $^{60}$Fe acts as an astrophysical clock on the $10^6$ year timescale, informing astrophysicists about the chemical history of our solar system.

At the time that the ERAWAST project was started, there was an ongoing controversy about the half-life of this interstellar clock isotope, which could only be resolved by a new measurement ERAWAST was able to provide sufficient $^{60}$Fe for the measurement [48,49] in addition to supplying $^{53}$Mn and $^{60}$Fe for neutron reaction studies, and enough $^{44}$Ti for radioactive beam studies at CERN and TRIUMF [50].

All in all, the ERAWAST collaboration was immensely successful at converting what would have been nuclear waste into some of the world's most valuable research material. The same approach at FRIB stands to deliver an even wider selection of short-supply isotopes that will fuel astrophysical research for years to come [51].

accounted for. This limits the amount of information that can be garnered from presolar grains where the ratios are known with a precision of a few percent. In order to rectify the models, a measurement of the neutron capture cross-section is required, which in turn requires access to a supply of $^{85}$Kr. During normal operations of FRIB, $^{85}$Kr will be created *en masse*, with a yield of up to $10^{12}$ atoms per second with the $^{86}$Kr primary beam. For a neutron capture experiment to be carried out at LANL's DANCE facility, a total of $6x10^{18}$ atoms are needed [38]. Since $^{85}$Kr decays slowly ($t_{1/2}$=10.7 y) krypton isotopes can be accumulated via isotope harvesting over the course of years to reach the necessary quantity, again without impacting normal operations or requiring dedicated beam-time at FRIB.

Similar processes can be undertaken for other branch-point isotopes because their characteristic lifetimes range from hours to tens of years, a timescale that is very well matched to what can be harvested at FRIB in commensal operation. As is proposed for $^{85}$Kr, the harvested isotopes will be taken to dedicated neutron beam facilities like DANCE to perform reaction measurements of neutron capture, $(n,\alpha)$, and (n,p) reactions. Recent advances in neutron intensity and detector capability have increased the range of measurements that are possible and decreased the amount of material needed for measurement. Calculations of required radioisotope quantities for a wide range of s-process scenarios have already been completed [38]. The neutron facilities and detectors are ready today if the right samples can be made available. In this way FRIB harvesting will provide a source of material that has otherwise proven highly elusive.

In addition to these measurements' impacts on *s*-process nucleosynthesis, they will also affect *r*-process nucleosynthesis studies. The typical *r*-process abundances that modelers attempt to match are based on differences between the observed abundances and what s-process models produce. As a result, s-process uncertainties propagate through to r-process scenarios. However, by combining with FRIB's fast and reaccelerated beam programs with isotope harvesting, researchers will gain a better handle on the origins of visible matter.

It is also important to note that the process of harvesting, purifying, and creating radioactive targets for nuclear astrophysics research will be similar, if not identical, to the procedure used to make targets for the stewardship science experiments mentioned above. In this way, the collective aims of both scientific communities will be able to draw upon shared expertise. Additionally, this will connect the SSP and nuclear astrophysics workforce pipelines by exposing young researchers to forefront research in both fields.

## Retaining National Expertise in Nuclear Science and Radiochemistry

Beyond excellent science, Isotope harvesting at FRIB also provides the nation with an opportunity to meet the growing need for trained nuclear and radiochemists. This is a critical component for all of the nuclear-related fields, from medicine to national security, because as the national need for isotopes grows, it is paralleled by demand for scientists trained to use and understand them.

The necessity for maintaining a well-educated workforce in radiochemistry in the US has been recognized for many years. In 2012 the National Academies Press published a National Research Council (NRC) report addressing the growing demands for- and limited supply of- trained radiochemists (*Assuring a future U.S.-based nuclear and radiochemistry expertise* [39]). The critical findings of the report are neatly summarized in the executive summary:

> *"The growing use of nuclear medicine, the potential expansion of nuclear power generation, and the urgent need to protect the nation against nuclear threats, to maintain our nuclear weapons stockpile, and to manage the nuclear wastes generated in past decades require a substantial, highly trained, and exceptionally talented workforce…*
>
> *…In order to avoid a gap in these critical areas, increases in student interest in these careers, in the research and educational capacities of universities and colleges, and sector specific on-the-job-training will be needed"*

Additionally, the recent NSAC long range plans for both the field of nuclear science overall and specifically for the DOE Isotope Program identify the need for a robust pipeline of highly trained scientists in order to ensure the progress in this field and in all related activities.

Owing to its position as a DOE facility at a major U.S. university and its numerous collaborations with other training centers, FRIB is in a strong position to meet many of the training recommendations of the report. The single best way to tap into this potential is to invest in the infrastructure needed for isotope harvesting. By giving students and postdoctoral trainees the opportunity for hands-on radiochemical research, the harvesting program will help ensure the existence of a well-trained radiochemistry workforce. It is also envisioned that this hands on training will attract visiting students and postdocs from other training centers in the nation for visits/internships at FRIB which would further enhance nuclear and radiochemistry expertise on a broader nationwide scale. This is especially true because of the wide range of applications that will be affected by isotope harvesting. The matching of radiochemical processing with applied science will demand that students become both technically apt and tuned-in to the national need. This pipeline of highly experienced radiochemists will be invaluable to national labs, hospitals, the energy sector, and academic research institutions across the US.

Finally, the development of a strong isotope harvesting program at FRIB will enable the distribution of isotopes to other centers which in turn will lead to trainees from outside of the traditional radiochemistry groups gaining expertise with working with radioactive materials to address specific scientific questions in other fields. While these trainees may not be experts in nuclear and radiochemistry, they will contribute to the scientific workforce with skills and knowledge in the uses of isotopes and isotopic techniques.

## Technology and Infrastructure for Accessing FRIB Isotopes

Fundamental to the motivation for harvesting isotopes from FRIB are the technical components of FRIB that are highly amenable for the process. It is not only the high rare-isotope creation rate that will enable the scientific opportunities described earlier, but also how and where the isotopes are created.

At the heart of FRIB is the super-conducting radiofrequency heavy-ion linear accelerator. The accelerator provides heavy-ion beams with powers up to 400 kW and energies up to 200 MeV/u for uranium, and higher energies for lighter ions. After acceleration, these beams impinge on a target where exotic nuclei (rare isotopes) are produced in-flight by nuclear reactions. A magnetic fragment separator is used to sort out isotopes of interest in multiple separation steps before transport into the FRIB beam line system. Early in the separation process the unreacted high-power primary beam is separated from the fragments and blocked in a rotating water-filled beam dump.

The water in the beam dump is primarily used to stop the beam, cool the beam dump, and carry the residual heat to heat exchangers. However in addition, when the heavy-ion beam enters the water the beam particles undergo multiple nuclear reactions such as fragmentation, fission, and spallation. Since the beam dump is meant to fully stop the beam, the creation of radioisotopes in the beam dump is expected to exceed creation in the in-flight target by an order of magnitude. Additionally, many of the long-lived beam dump radioisotopes, like $^{47}Ca$, $^{67}Cu$ and $^{225}Ra$, will come to rest as aqueous ions in the flowing water where they can be readily transported and extracted, *i.e.* “harvested”. This harvesting process could therefore occur completely independent of the FRIB operational mission of providing rare isotope beams for basic science. As seen in Figure 6, the FRIB baseline is very well-suited for providing access to the shortest-lived and rarest isotopes. The additional capabilities garnered by isotope harvesting efficiently extend FRIB’s reach to allow accumulation of isotopes with longer half-lives for offline experiments.

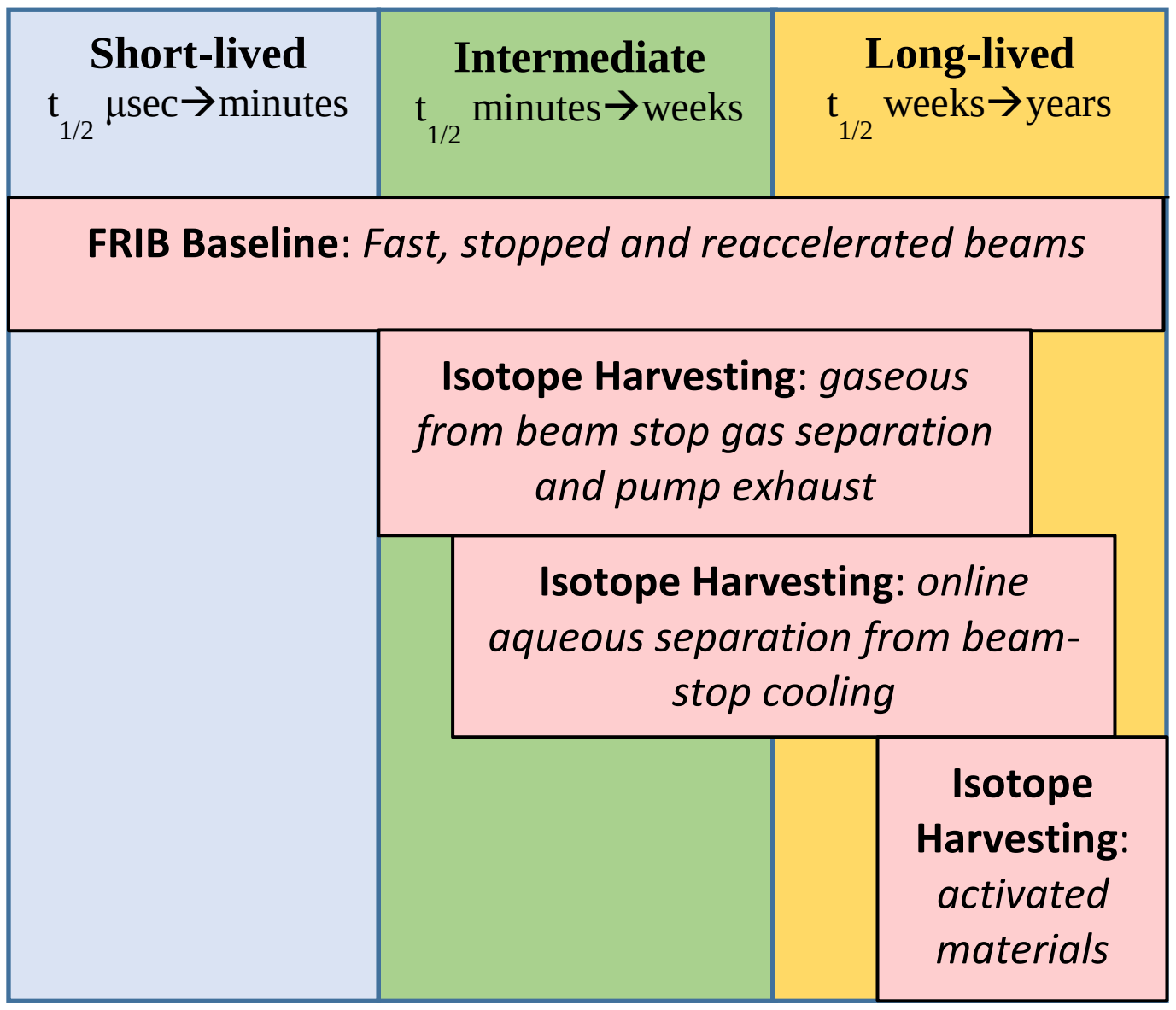

*Figure 6: An overview of the rare-isotope opportunities at FRIB, in terms of methodology and half-life.*

The water-cooled beam dump is not the only location where useful isotopes will be created. As all other devices that block or collimate the beam will also act as generation points for radionuclides, there are multiple locations to tap for harvesting. In particular, the beam-line pump exhaust will entrain many gaseous radionuclides that are created via collimation and charge selection. When the beam interacts with a collimator, natural beam-heating will allow induced gaseous radioisotopes (like noble gasses) to diffuse into the beam-line and eventually into the vacuum pump exhaust. By flushing the pump exhaust with inert gas, the valuable noble gas isotopes like $^{211}Rn$ and $^{76}Kr$ can be directed to collection traps like the Metal-Organic-Frameworks (MOFs) described in the sidebar.

Other beam stops and slits that are not readily accessible through water cooling or gaseous diffusion will also become activated. These will slowly accumulate long-lived radioisotopes like $^{60}Fe$ and $^{32}Si$ over the course of routine beam operations. Given the proper infrastructure, these materials could be post-processed to recover the highly valuable long-lived isotopes from these locations as well.

These three opportunities: The primary beam dump; the pump exhaust system; and the activated components, in addition to other collection points like fragment catchers and water-filled intercepting devices will enable harvesting of isotopes with a wide range of half-lives and chemistries. Since harvesting is an accumulating process, and because isotopes will be accessed from areas where the isotope creation rates are very high, relevant quantities of radionuclides will be made available for offline experiments through the harvesting program.

### Technology for Harvesting: Membrane Contactors

One exciting recent development in separations technology is the membrane contactor. Membrane contactors allow constant countercurrent extraction of ions and gases across a hollow fiber-supported membrane. Depending on the characteristics of the membrane, these devices can be made chemically specific, allowing fine-tuning of the extraction process. For harvesting at FRIB the membrane contactor is an important advancement for two reasons, first because there will be such a wide array of isotopes to parse, and second because it will allow radionuclides to be harvested from the primary cooling flow using a mobile secondary stream. The secondary stream can be transported to other locations in the lab without actually transferring the primary cooling water out of the target facility. This option is non-invasive to FRIB operation, as it will transport valuable radionuclides without interruption.

## Needed infrastructure

At FRIB, the beam dump cooling water loop and the pump exhaust handling systems are already in place inside the non-conventional utilities (NCU) area of the target hall. These systems will be tapped as part of the isotope harvesting project. The cooling water loop is already fitted with ports for a secondary purification loop outside of the target facility. These ports are accessible in a small room that will be repurposed for isotope harvesting, termed the “harvesting room”. The pump exhaust system will be tapped in a similar fashion: a slow flow of inert gas will drive the pump exhaust through tubing into the harvesting room. There, the rapidly advancing technologies of membrane contactors and metal organic frameworks (MOFs) will be used to collect long-lived radon, krypton, and xenon isotopes (see sidebars).

In order to handle the accumulating radioactivity, the harvesting room will require a shielded cell. Inside the cell, the membrane contactors, MOFs, and ion exchangers will be housed for collection. Isotopes will be collected with these, inside of moveable shielded containers. After collection, the containers will be loaded onto carts and taken out of the target hall for processing. A new radiochemistry facility, termed the Isotope Processing Area (IPA), depicted in figure 7, is envisioned for the processing. It can be accommodated in a new experimental hall at FRIB that is in the schematic design stage. The hall is planned to have sufficient space and appropriate infrastructure to house both the IPA and a new high rigidity spectrometer (HRS), which has been advocated by the FRIB users' community, and is currently receiving research and development funding.

Figure 7 shows a schematic layout of the IPA, which will house separations equipment needed to perform the chemical purifications on harvested isotopes. The essential pieces of the infrastructure are shielded cells with dedicated ventilation, shielded fume hoods for preparation of samples, space for short-term waste storage, and analytical equipment. The operational concept will be to collect the harvested isotopes in the harvesting room, transfer them to the IPA, and finally to purify them for use in on-site and off-site applications. The aim of processing will be to produce a specific radioisotope in a simple chemical form, most often as a trapped gas or as a simple salt dissolved in aqueous solution. Further chemical processing or radiolabeling will take place at the Radiochemistry Laboratory at MSU or at an off-site location as needed by the user community.

The functionality of the IPA will be split to match the three physical forms the harvested isotopes will arrive in: gaseous from the pump exhaust and membrane contactors; in aqueous resins from the cooling water loop; or trapped inside of a solid piece of activated equipment. This will be achieved by having three sets of dual shielded cells, with each hot cell dedicated to a single form. Since the palette of harvested isotopes will be different for every beam that is used, the

### Harvesting Technology: Metal Organic Frameworks (MOFs)

A promising new technology for krypton, xenon, or radon harvesting is an adsorption-based process using selective, solid-state adsorbents called metal-organic-frameworks (MOFs). An important advantage of MOFs is their chemical tenability, as MOFs can be tailor-made for optimal selectivity in capturing Kr, Xe, or Rn at room temperatures.

D. Banerjee and coworkers at Pacific Northwest National Lab have recently synthetized a new MOF with a pore size specifically tuned to adsorb xenon [52]. Preliminary tests have shown that this material has superior properties for xenon adsorption in terms of efficiency, selectivity, and capacity, and can operate within a diluted gas stream.

MOFs have higher efficiency, selectivity, and capacity at room temperature over current xenon adsorbents like activated charcoal and Ag-loaded zeolites at cryogenic temperatures. In addition, MOFs require limited pre-treatment of the intake gases and no cryogenic operation. Collection systems will be lightweight, and suitable for low power and space-limited deployment.

This novel technology is a perfect fit for FRIB harvesting because a large portion of the off-gas stream from the FRIB beam dump will be diluted with nitrogen, and cryogenic treatment is not feasible. MOFs will allow efficient online trapping at FRIB without interference.

cells will allow processing of one set of isotopes from a previously harvested beam while preparing for a different set of samples from the current beam. Cross contamination between isotopes in a single hot cell will be prevented by processing in campaigns with a clean-out in between campaigns, and by decay.

While the physical space for operating an isotope harvesting program is currently in the schematic design phase, additional funds are needed, specifically for the following capital equipment:

- Radioisotope-preparation shielded cells (with separate functionalities for activated aqueous and component processing)
- Shielded radiochemistry fume hoods and a glove box (for gas-phase processing, performing secondary separations and radio-analytical prep work)
- Sealed work-surfaces for radiochemical handling
- Laboratory ventilation for open radiochemical work (with active filtration and monitoring), sustained negative pressure with respect to the surrounding facility (*i.e.* an enclosed laboratory with air-locked and monitored entranceways).
- Laboratory security, monitoring, and protective equipment (including mobile and modular shielding, interlocked entranceways, active dose monitors, and hand and foot monitors at entranceways).
- Analytical equipment: radio-HPLC, MP-AES, autoradiography, liquid scintillation, HPGE, ionization chambers, alpha counting system
- Short-term activated waste storage, including liquid waste storage equipment, and radioactive-transport casks and carrier equipment

Creating a fully-operational, safe, and secure laboratory with this capital equipment will provide access to some of the most exotic long-lived nuclei ever created. This facility will act as the interface between the applied-science User community and the FRIB project, and will promote high impact science across multiple disciplines. Additionally, the buildup of the IPA can be sequenced to match the startup phases of FRIB, where harvesting infrastructure is added in a phased-build plan. This phased approach will allow the processing techniques at the IPA to come online in-step with FRIB's march to full power- increasing the radionuclidic and chemical purity of products as the isotope creation capacity increases.

## A phased approach for implementing isotope harvesting at FRIB

Construction of the Isotope Harvesting Facility will be carried out in four, year-long phases from FY 2020 to FY 2023.

The first phase will focus on civil construction of the ventilated workspace (IPA) with three shielded radioisotope fume hoods, and installation of a shielded cell adjacent to the target facility in the harvesting room. The harvesting room cells will have access to the cooling water and gas streams. Based upon the design of the water cooling system at FRIB, there will be an opportunity (already at the completion of the first phase) to draw limited quantities of radioisotopes by routing a small fraction of the streams to an access point. This shielded access will allow us to readily monitor the water quality and radioactivity transport in the NCU. The samples will be minimally processed in the shielded fume hoods, and taken to the MSU radiochemistry laboratory for analysis. The operations in the first phase will serve to validate the extraction techniques, evaluate their performance, and provide an important stepping stone to access larger quantities of by-product isotopes.

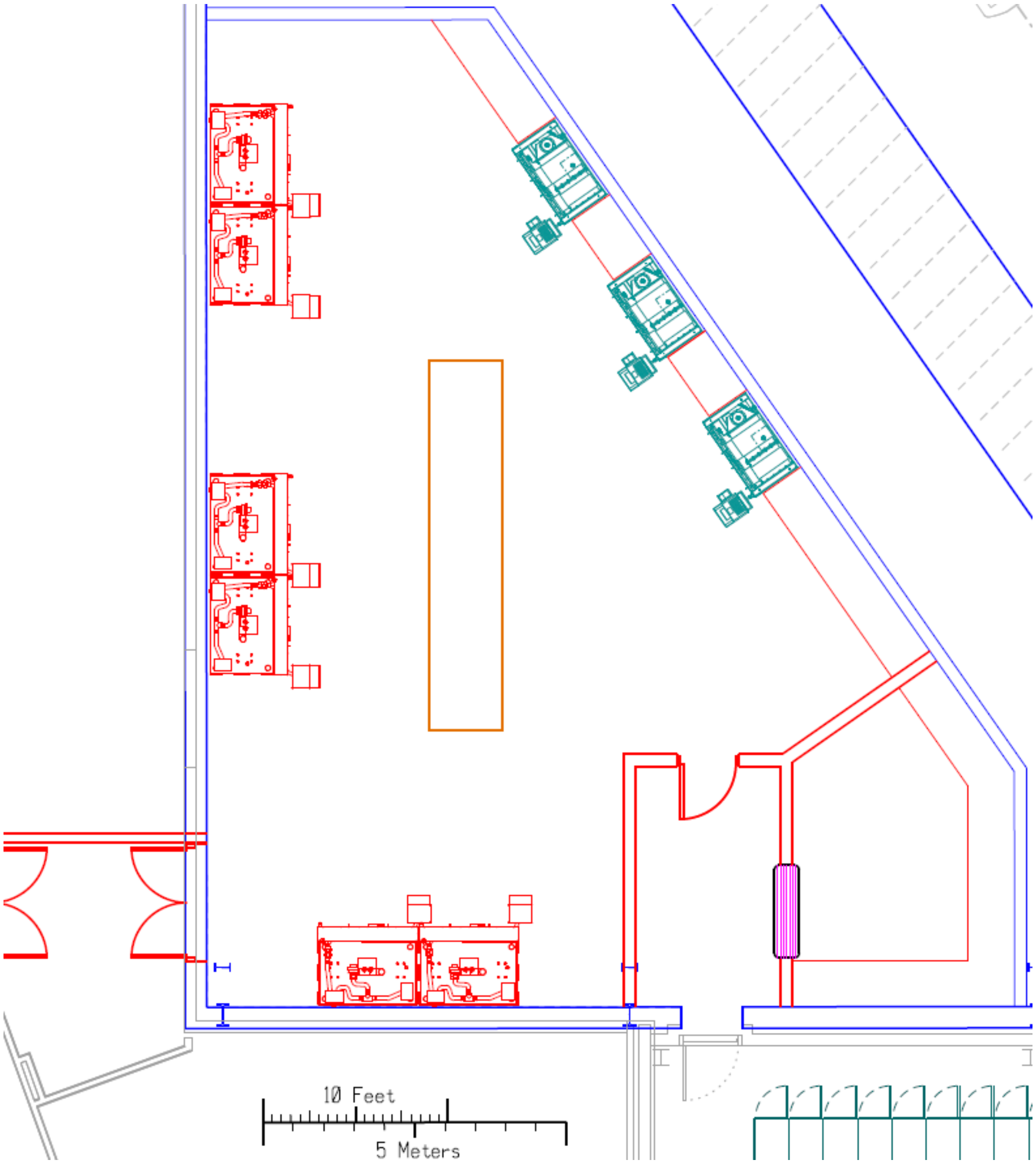

*Figure 7: A proposed layout for the Isotope Processing Area (IPA) at FRIB. The layout includes shielded cells (red), shielded hoods (green), air-locked entrances and a packaging area.*

The second phase of building will encompass the installation of one set of dual-shielded cells, and the addition of analytical equipment in-house at the IPA. The analytical equipment will include a liquid scintillation counter, high purity germanium detectors, and alpha counting equipment. During this phase the goal of the IPA will be to begin making samples available for use on a small scale, mostly for the purpose of evaluation. The main development goal during this period will be to maximize the radionuclidic purity of samples.

Phase three will rely on the high radionuclidic purity techniques established during phase two to shift focus to radiochemical purity, high specific activity, and test reactions. Work during phase three will be the installation of the final two shielded-cell pairs and the inclusion of additional analytical equipment intended for the evaluation of the chemical purity and specific activity, including radio-HPLC, radio-TLC, radio-GC, autoradiography equipment, and a microwave-plasma atomic-emission spectrometer. During this time, as dictated by the quality and reproducibility of extraction and purification techniques, the fractional flows of NCU water and gases will be increased to allow access to higher quantities of isotopes.

Phase four will culminate the construction by installation of a packaging and distribution room with a pass-through to the ventilation interlock. This final addition to the facility will allow the harvested and processed isotopes (of assured and reproducible quality) to be shipped routinely to outside users.

At the end of the phased construction, the gas, water, and activated materials radiochemical processing units will be operating according to the particular needs of the user community. The sets of dual shielded cells will allow workup of harvested isotopes from one primary beam while preparing for processing another batch from the next beam. After the main radioisotopic components are parsed in the shielded cells, finer tuned purifications will be employed in the radioisotope fume hoods. After purification, analytical protocols will assure the quality and purity of the isotopes.

By following this four-phased approach, the investment in isotope harvesting at FRIB will follow its development from a novel approach to accessing rare isotopes to a fully operational and high-quality source of rare isotopes. As described in the preceding sections, there is an established need for radioisotopes in a multitude of research settings. Isotope harvesting at FRIB will provide the link between the research community and the unique opportunity provided by the Facility for Rare Isotope Beams.

Table 3. An outline of the phased-build approach for isotope harvesting at FRIB

| | **Total Equipment Costs** | **pre-ops** | **Major Equipment** | **Year-End Objective** |
|---|---|---|---|---|
| | **(M$)** | **(M$)** | | |
| **FY 2020** | 2.5 | 0 | ventilation, shielded hoods, monitoring equipment | demonstration of safe operations and harvesting |
| **FY 2021** | 3 | 0.5 | shielded cells, radioanalytical instruments | high radionuclidic purity harvesting |
| **FY 2022** | 2 | 0.5 | shielded cells, analytical chemistry instruments | high chemical purity harvesting |
| **FY 2023** | 0.5 | 0.5 | packaging equipment | fully operational harvesting, with isotope distribution of research quantities of isotopes to users |

## Summary and Conclusion

Isotope harvesting at FRIB taps into the unique isotope creation capabilities of one of the most sophisticated nuclear physics facilities in the world. The opportunity afforded by investment in isotope harvesting infrastructure is unparalleled and should be prioritized. The appropriate time to implement harvesting is during the early stages of FRIB operation, as the facility builds up to full power. The harvesting program will have impact in multiple fields, including medicine, physiology, basic nuclear science, energy, horticulture and astrophysics. The investment required to access the valuable isotopes that will be formed at FRIB will lead to advances in both technology and in our understanding of the world.

# Appendix 1. A table of radionuclides to be harvested at FRIB

Table A1 gives a list of radionuclides already identified as important harvesting targets, and their expected production rates in the FRIB beam dump. The production rate includes formation of the isotope as a result of direct production in the beam and by parent isotope decay using the LISE++ utility [40].

The table is organized by primary beam, and rates are given with units appropriate to the half-life of the radionuclide of interest. For example, isotopes with half-lives on the order of weeks to years, and beyond, are given in mCi/wk, and those with half-lives on the order of hours and days are typically given in mCi/day. Three isotopes are treated slightly differently, $^{221}$Rn, $^{223}$Rn and $^{77}$Kr, because of their short half-lives. The units for $^{221}$Rn and $^{223}$Rn are given as a number of atoms in steady state, and $^{77}$Kr is given in mCi/hr. Typical operation at FRIB will run in beam 'campaigns' where a single primary beam will be run for a time period usually not less than one week. Therefore production rates given as daily or weekly yields are relevant for planning purposes for daily or weekly harvesting campaigns.

The numbers in the table do not reflect any losses from chemical processing, which will be dependent on the exact harvesting procedure. It should be noted, however, that all of the isotopes listed have a reasonable pathway for chemical extraction and purification, assuming that the extraction process will not take more than a few minutes for noble gases, and not more than a few hours for ions.

Additionally, all isotopes listed have a pathway to high radionuclidic purity where the recognized application demands it (e.g. in medicine), or high isotopic purity (*e.g.* for some stewardship and astrophysics applications). This is why the harvested isotope is often a generator parent of an important radionuclide. For example, extraction of $^{47}$Ca during $^{48}$Ca irradiations will lead to high-radionuclidic purity $^{47}$Sc for medical use, whereas direct harvesting of $^{47}$Sc will unavoidably result in co-harvesting of scandium contaminants like $^{46}$Sc and $^{48}$Sc which would preclude medical applications. Note: some applications, such as EDM experiments, are not hindered by isotopic impurities, and therefore careful parent-daughter extractions are not needed. For example, $^{221}$Rn and $^{223}$Rn cannot be separated from each other, or from other radon isotopes. However probing these atoms for nuclear EDMs involves exciting an isotopically specific atomic transition, meaning that co-harvesting of radon isotopes will not interfere with a measurement. The specific case of how to obtain high purity when harvesting $^{211}$Rn for production of $^{211}$At is treated in Appendix 2. In principle, all of the isotopes listed in Table A1 will be able to be obtained using similar techniques.

Finally, many of the isotopes in Table A1 will be created from multiple beams, not just the primary beam listed. Therefore, there will be additional quantities of these isotopes available for harvesting when a different primary beam is planned.

| Element | Mass # | $t_{1/2}$ | Research Fields | Use | Beam | production rate | (units) |
|---|---|---|---|---|---|---|---|
| Mg | 28 | 21 h | plants | $Mg^{2+}$ transport tracer | $^{36}S$ | 2.1E+03 | mCi/d |
| Si | 32 | 153 y | environmental tracing | silica cycle | $^{36}S$ | 1.1E+00 | mCi/wk |
| Ca | 47 | 4.5 d | medicine | generator of therapeutic $^{47}Sc$ | $^{48}Ca$ | 1.0E+04 | mCi/d |
| Ti | 44 | 60 y | astrophysics, medicine | supernova observable, generator for $^{44}Sc$ | $^{58}Ni$ | 2.5E-01 | mCi/wk |
| V | 48 | 16 d | stewardship, plants | neutron reaction networks, nitrogen fixation | $^{58}Ni$ | 2.6E+03 | mCi/wk |
| Cr | 48 | 22 h | stewardship | generator of $^{48}V$ without $^{49}V$ impurity | $^{58}Ni$ | 6.1E+02 | mCi/d |
| Fe | 52 | 8.3 h | plants, medicine | Fe – PET, micronutrient tracing, $^{52m}Mn$ generator | $^{58}Ni$ | 9.8E+02 | mCi/d |
| Fe | 60 | 1.5 My | astrophysics | s-process branch data, neutron capture | $^{64}Ni$ | 2.8E-05 | mCi/wk |
| Ni | 63 | 101 y | RTG | RTG | $^{64}Ni$ | 5.6E+00 | mCi/wk |
| Zn | 62 | 9.2 h | materials, biochemistry | PAC Copper, PET, $^{62}Cu$ generator | $^{64}Zn$ | 3.2E+03 | mCi/d |
| Se | 72 | 8.5 d | medicine | generator of $^{72}As$ for PET | $^{74}Se$ | 2.2E+03 | mCi/wk |
| Se | 73 | 7.2 h | materials, biochemistry | PAC Arsenic | $^{74}Se$ | 1.6E+04 | mCi/d |
| Cu | 67 | 2.6 d | medicine | beta immunotherapy, theranostic match to $^{64}Cu$ | $^{76}Ge$ | 6.9E+02 | mCi/d |
| As | 77 | 35 h | astrophysics | neutron TOF | $^{76}Ge$ | 9.6E+03 | mCi/d |
| Ge | 68 | 271 d | medicine | generator for $^{68}Ga$ | $^{78}Kr$ | 1.6E+01 | mCi/wk |
| Kr | 76 | 14.8 h | medicine | generation of PET halogen $^{76}Br$, PAC | $^{78}Kr$ | 2.5E+03 | mCi/d |
| Kr | 77 | 74 m | materials, biochemistry | PAC Bromine, Parent to $^{77}Br$, therapy | $^{78}Kr$ | 6.7E+03 | mCi/hr |
| Sr | 82 | 25 d | medicine | cardiac PET parent for $^{82}Rb$ | $^{92}Mo$ | 7.9E+02 | mCi/wk |
| Zr | 86 | 16.5 h | materials, biochemistry | PAC Yttrium, generator of $^{86}Y$ PET | $^{92}Mo$ | 2.9E+03 | mCi/d |
| Zr | 88 | 83 d | stewardship | neutron reaction networks | $^{92}Mo$ | 6.3E+02 | mCi/wk |
| Mo | 90 | 5.7 h | plants | Mo - PET nitrogen fixation enzymes | $^{92}Mo$ | 2.8E+03 | mCi/d |
| Pd | 100 | 3.7 d | medicine | Pd nanoparticles, delayed PET via $^{100}Rh$, PAC | $^{106}Cd$ | 6.5E+02 | mCi/d |
| Pd | 103 | 17 d | medicine | generator of Auger emitter $^{103m}Rh$ | $^{106}Cd$ | 2.7E+03 | mCi/wk |
| In | 111 | 2.8 d | materials, biochemistry | PAC, SPECT | $^{112}Sn$ | 2.9E+03 | mCi/d |
| Te | 119 | 4.7 d | medicine | generator for Auger emitter $^{119}Sb$ | $^{124}Xe$ | 3.0E+03 | mCi/wk |
| Xe | 122 | 20 h | medicine | parent to short-lived PET $^{122}I$ | $^{124}Xe$ | 2.8E+03 | mCi/d |
| Ba | 128 | 2.4 d | medicine | generator for short-lived alkali PET | $^{136}Ce$ | 6.8E+02 | mCi/d |
| Ce | 134 | 3.2 d | medicine | generator for short-lived lanthanide PET | $^{136}Ce$ | 4.2E+03 | mCi/wk |
| Nd | 140 | 3.4 d | medicine | generator for short-lived lanthanide PET | $^{144}Sm$ | 4.1E+03 | mCi/wk |
| Nd | 139m | 4.4 h | materials, biochemistry | PAC praseodymium, PET parent | $^{144}Sm$ | 8.6E+02 | mCi/d |
| Ce | 143 | 33 h | materials, biochemistry | PAC praseodymium | $^{150}Nd$ | 1.8E+02 | mCi/d |
| Pm | 147 | 2.6 y | chemistry, RTG | the chemistry of Pm, RTG | $^{150}Nd$ | 2.6E+01 | mCi/wk |
| Gd | 146 | 48 d | medicine | parent to positron emitter $^{146}Eu$ | $^{156}Dy$ | 1.2E+02 | mCi/wk |
| Gd | 147 | 38 h | materials, biochemistry | PAC promethium | $^{156}Dy$ | 4.6E+02 | mCi/d |
| Gd | 148 | 71 y | batteries | alpha lanthanide RTGs | $^{156}Dy$ | 4.7E-01 | mCi/wk |
| Gd | 149 | 9.3 d | stewardship | parent to $^{149}Eu$- neutron network | $^{156}Dy$ | 1.2E+03 | mCi/wk |
| Eu | 150 | 36 y | stewardship | neutron network | $^{156}Dy$ | 6.0E-02 | mCi/wk |
| Hf | 172 | 1.9 y | chemistry | parent to $^{172}Lu$ for PAC | $^{174}Hf$ | 2.1E+01 | mCi/wk |
| Hf | 173 | 24 h | stewardship | parent to $^{173}Lu$: neutron network | $^{174}Hf$ | 1.8E+03 | mCi/d |
| Tm | 168 | 93 d | stewardship | neutron network measurements | $^{176}Yb$ | 5.8E+01 | mCi/wk |
| Lu | 174 | 3.3 y | stewardship | neutron network measurements | $^{186}W$ | 6.5E-01 | mCi/wk |
| Tl | 204 | 3.8 y | RTG | Beta RTG | $^{208}Pb$ | 1.4E+00 | mCi/wk |
| Sr | 91 | 9.7 h | astrophysics | Parent to $^{91}Y$ for NTOF | $^{238}U$ | 9.1E+02 | mCi/d |
| Zr | 95 | 64 d | stewardship, astrophysics | neutron network | $^{238}U$ | 1.1E+02 | mCi/wk |
| Mo | 99 | 66 h | materials, biochemistry | PAC Tc, SPECT | $^{238}U$ | 3.0E+02 | mCi/d |
| Ru | 103 | 39 d | medicine | parent to Auger emitter 103mRh | $^{238}U$ | 1.9E+02 | mCi/wk |
| Ru | 105 | 4.4 h | astrophysics | parent to $^{105}Rh$, neutron activation | $^{238}U$ | 1.6E+03 | mCi/d |
| Ru | 106 | 372 d | astrophysics | NTOF target | $^{238}U$ | 2.1E+01 | mCi/wk |
| Ba | 140 | 12.7d | astrophysics | parent to $^{140}La$ | $^{238}U$ | 1.5E+02 | mCi/wk |
| Rn | 211 | 14.6 h | medicine | generation of alpha therapeutic $^{211}At$ | $^{238}U$ | 4.3E+02 | mCi/d |
| Rn | 221 | 25 m | EDM | EDM | $^{238}U$ | 2.1E+11 p | steady state |
| Rn | 223 | 23 m | EDM | EDM | $^{238}U$ | 5.7E+10 p | steady state |
| Ra | 225 | 15 d | EDM | EDM | $^{238}U$ | 4.9E+00 | mCi/wk |
| Ac | 225 | 10 d | medicine | generator for $^{213}Bi$, or direct alpha therapy | $^{238}U$ | 4.4E+01 | mCi/wk |
| Ac | 227 | 21.7 y | medicine | impurity in $^{225}Ac$ / parent to $^{227}Th$ | $^{238}U$ | 3.4E-02 | mCi/wk |
| Th | 227 | 18.7 d | medicine | generator for $^{223}Ra$ | $^{238}U$ | 6.4E+01 | mCi/wk |
| Th | 228 | 1.9 y | medicine | generator $^{212}Pb/^{212}Bi$ | $^{238}U$ | 8.1E+00 | mCi/wk |
| Pa | 229 | 1.5 d | EDM | level splitting, octupole deformation, EDM | $^{238}U$ | 3.9E+02 | mCi/d |
| Th | 229 | 7.9 ky | medicine, EDM | nuclear clock, $^{225}Ra$ parent, $^{225}Ac$ parent | $^{238}U$ | 2.0E-03 | mCi/wk |

# Appendix 2. Attaining isotopic purity from mixed radioisotopes

While the enormous isotope creation capacity at FRIB is highly beneficial, in many cases the co-creation of so many radionuclides could potentially lead to low isotopic purity of the desired radionuclide. This can be overcome by either taking advantage of isotope 'generation', or by careful timing, or both. Therefore, although FRIB will provide a complicated mix of radioisotopes, it is still feasible to obtain pure isotopes in high quantities.

The case of $^{211}$At provides a useful illustration of the pathway to isotopic purity in the face of a multi-isotopic source. $^{211}$At is a high-priority isotope for alpha-radioimmunotherapy and other targeted internal therapies. It will be created directly at FRIB at a high rate, about $10^9$ atoms per second. Also, all other At isotopes, from mass 195-212 will co-created at a similar or higher rate. Clearly, by waiting a few hours after harvesting astatine isotopes from FRIB, the shorter-lived isotopes will simply decay away, and their daughters can be chemically removed. However $^{207, 208, 209}$At and especially $^{210}$At cannot be removed efficiently by waiting, because their half-lives are too similar to the 7 hour half-life of $^{211}$At. Additionally, $^{210}$At decays to $^{210}$Po, which is highly toxic: thereby prohibiting medical administrations of $^{211}$At with high $^{210}$At impurity. Therefore, it is impossible to obtain $^{211}$At for medicine from FRIB by chemically harvesting astatine directly.

However, there is a very useful way to solve the problem of $^{211}$At purity from FRIB: through a generator route. As described earlier in this whitepaper, $^{211}$At is generated by the decay of $^{211}$Rn, which has a half-life of 15 hours. $^{211}$Rn is predicted to be created in the FRIB beam dump at a rate of 430 mCi/d (or about $2 \times 10^{10}$ atoms per second) while the full power $^{238}$U beam is being used. Additionally, since $^{211}$Rn is a noble gas, it can be extracted from the cooling water across a liquid-gas membrane contactor, meaning that it is rapidly available during operations without complicated wet-chemical processing.

Just as with harvesting astatine, collecting radon isotopes will be a low isotopic purity endeavor, as many Rn isotopes are created simultaneously. However, during $^{238}$U irradiation, only four isotopes of Rn will be created with a half-life longer than 1 hour, and of those four, only two have a decay branch to long-lived astatines (211, and 210). Taken together these two At parents would also lead to a $^{210}$Po generation issue, but this problem is easily overcome by taking advantage of the different half-lives of $^{211}$Rn and $^{210}$Rn (14.6 h, and 2.4 h respectively), and their different branching ratios to astatines (73% and 4% respectively). Therefore, one simple protocol for obtaining pure $^{211}$At is to harvest all Rn isotopes from a membrane contactor for 24 hours, and then hold them to decay for 24 hours (see figure A2). At this time there will be both $^{210}$At and $^{211}$At present, which is not useful. However if, at the end of the 24 hour decay period, the radon is transferred to a new gas trap (leaving astatines behind), the astatine generation will begin afresh, only now with a much lower quantity of $^{210}$Rn. After one more day, $^{211}$At will have formed in the new trap with >99.99% isotopic purity.

As an added benefit, the $^{211}$Rn generator can be shipped to a remote medical center and create $^{211}$At *en route*. One day after the gas trap transfer, 60 mCi of $^{211}$At will be available for use in radiopharmaceutical research. After another day, the generator can be milked again to obtain 19 mCi additional $^{211}$At. In principle, this process can occur every single day that $^{238}$U beam is on target, making FRIB unmatched in its capacity to create $^{211}$At. If this capacity is coupled to the *University Network for Accelerator Production of Isotopes*, it will provide an important source of $^{211}$At in support of clinical trials across the US.

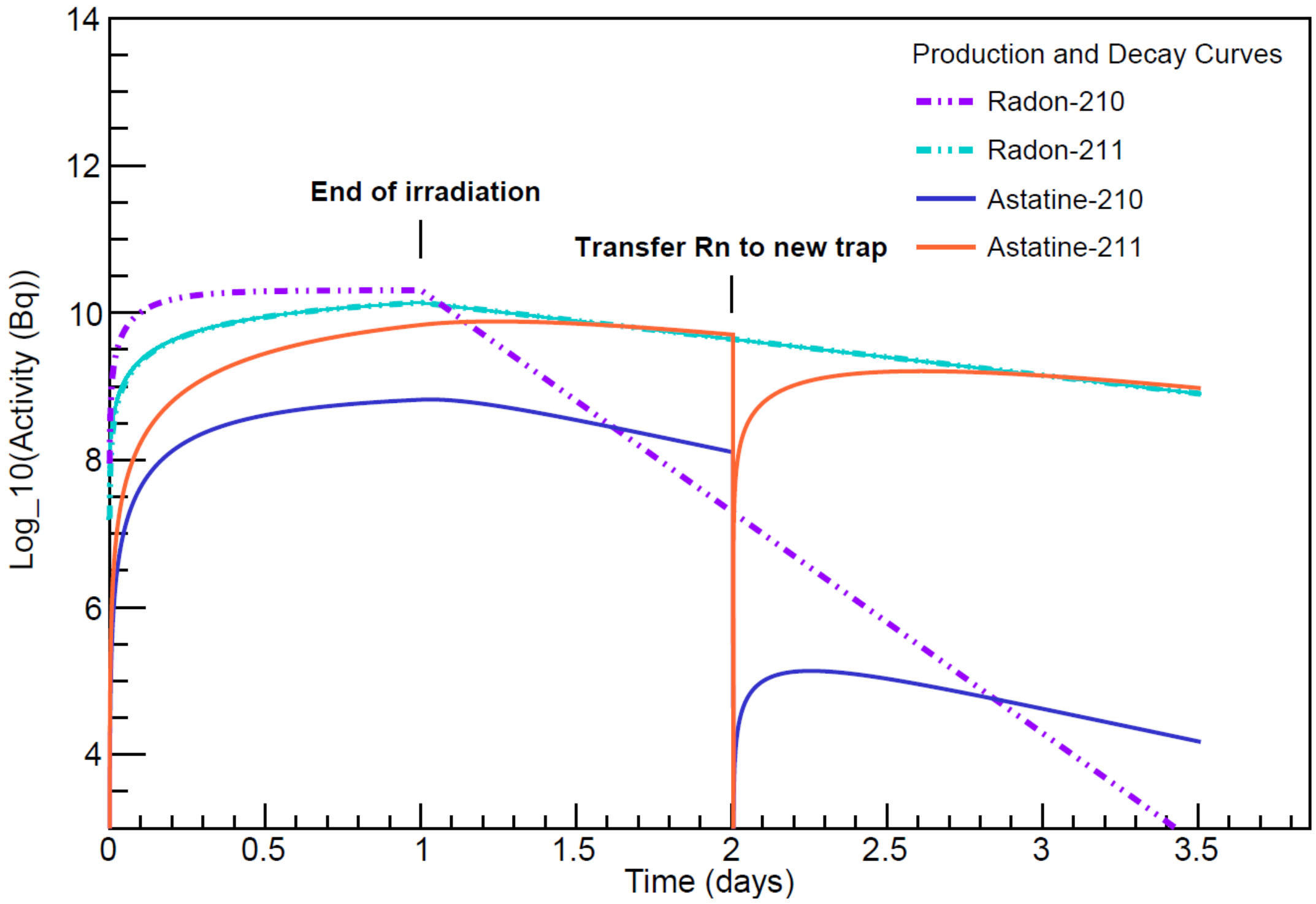

*Figure A2: Radioisotopic activities in the $^{211}$Rn-based scheme for obtaining high quantities of isotopically pure $^{211}$At from FRIB. The activity of the relevant Rn isotopes, $^{210}$Rn and $^{211}$Rn are shown with hashed lines, and the activity of their astatine daughters, $^{210}$At and $^{211}$At, are shown with solid lines. In this scheme, Rn isotopes are collected for 24 hours, from the membrane contactor and off-gas system at FRIB (cessation of collection is marked as "end of irradiation" on the plot). At this time, the $^{210}$Rn and $^{211}$Rn activities begin to diverge due to their different half-lives. After another day, the Rn isotopes are transferred to a new gas trap (marked on plot). Following another 24 hours (day 3 on the plot), $^{211}$At has a radioisotopic purity over 99.99%, in excess of 50 mCi.*

## List of acronyms

AGB- Asymptotic Giant Branch
CERN-ISOLDE- Center for European Nuclear Research, Isotope Separator-Online
DANCE- Detector for Advanced Neutron Capture Experiments
DARPA- Defense Advanced Research Projects Agency
DOE- Department of Energy
EDM- Electric Dipole Moment
FRIB- Facility for Rare Isotope Beams
IDPRA- Isotope Development and Production for Research and Applications
IPA- Isotope Processing Area
MDM- Magnetic Dipole Moment
MEDICIS- Medical Isotopes from CERN-ISOLDE
MEMS- Micro-Electro-Mechanical Systems
MOFs- Metal Organic Frameworks
NIDC- National Isotope Distribution Center
NNSA- National Nuclear Security Administration
NSAC- Nuclear Science Advisory Committee
NSAC-I - Nuclear Science Advisory Committee Isotopes Subcommittee
PAC- Perturbed Angular Correlation
PET- Positron Emission Tomography
PNI-Polarized Nuclear Imaging
RTG- Radio-thermal Generator
SPECT- Single Photon Emission Computed Tomography
SSP- Stockpile Stewardship Program